\documentclass[aps,twocolumn,showpacs]{revtex4}
\usepackage{graphics}

\newcommand\rx{x}
\newcommand\ry{y}
\newcommand\ra{a}
\newcommand\rb{b}

\newcommand\eref[1]{(\ref{#1})}

\begin{document}

\title
{
  Polymer models with competing collapse interactions on Husimi
  and Bethe lattices
}

\author{M. Pretti}

\affiliation {
  Consiglio Nazionale delle Ricerche - Istituto dei Sistemi Complessi (CNR-ISC), \\
  Dipartimento di Scienze Applicate e Tecnologia (DISAT), Politecnico di Torino,
  Corso Duca degli Abruzzi 24, I-10129 Torino, Italy
}

\date{\today}

\begin{abstract}

In the framework of Husimi and Bethe lattices, we investigate a
generalized polymer model that incorporates as special cases
different models previously studied in the literature, namely,
the standard \emph{interacting self-avoiding walk}, the
\emph{interacting self-avoiding trail}, and the
\emph{vertex-interacting self-avoiding walk}. These models are
characterized by different microscopic interactions, giving
rise, in the two-dimensional case, to collapse transitions of
an apparently different nature. We expect that our results,
even though of a mean-field type, could provide some useful
information to elucidate the role of such different \emph{theta
points} in the polymer phase diagram. These issues are at the
core of a long-standing unresolved debate.

\end{abstract}


\maketitle

\section{Introduction}

The collapse transition of a polymer chain in dilute solution
is one of the most classical topics in polymer
physics~\cite{deGennes1979,DesCloizeauxJannink1990,Vanderzande1998}.
Such a phenomenon arises from a competition between excluded
volume and some kind of attractive interaction of the monomers
with one another. Depending on which of the two effects
prevails, the polymer takes on a swollen state (coil) or a
compact one (globule), also denoted as good- or bad-solvent
regimes, respectively. The two regimes are separated by a phase
transition (driven by temperature, or by some actual change in
the solvent quality), which is characterized by specific
properties of the polymer, usually denoted as the \emph{theta}
state.

Different lattice models have been proposed to investigate
these phenomena. In the standard \emph{interacting
self-avoiding walk} (ISAW)
model~\cite{deGennes1979,DesCloizeauxJannink1990,Vanderzande1998},
polymers are forbidden from visiting a lattice site more than
once, and a contact interaction is assigned to nearest-neighbor
sites visited by nonconsecutive monomers. Alternative models
are the \emph{interacting self-avoiding trail}
(ISAT)~\cite{MassihMoore1975,Foster2009} and the
\emph{vertex-interacting self-avoiding walk}
(VISAW)~\cite{BloteNienhuis1989,FosterPinettes2003}, both
characterized by the fact that only lattice bonds (not sites)
are subject to the single-visit constraint. The two models
differ in the fact that the ISAT is allowed to cross itself,
whilst the VISAW is not. In both cases, the self-attractive
interaction is associated with the multiply-visited sites.

According to the paradigm of universality, one would expect
that the critical behavior of all three of the above models
should be described by the same universality class. Indeed,
this seems to be the case in the good-solvent regime (which is
itself a critical state), but there are several evidences that,
in two dimensions, the theta state belongs to three different
universality classes and, for the ISAT and VISAW cases, it is
also associated to the onset of a peculiar
\emph{maximally-dense}
phase~\cite{FosterPinettes2003,Foster2009,BediniOwczarekPrellberg2013_PRE}
and other more subtle
features~\cite{NahumSernaSomozaOrtuno2013,VernierJacobsenSaleur2015}.
Note that two-dimensional models have attracted great
theoretical interest, due to the availability of rigorous
results coming from exact solutions and/or conformal field
theories, through the analogy with ``magnetic'' $O(n)$ models
in the limit ${n \to
0}$~\cite{deGennes1979,DesCloizeauxJannink1990,Vanderzande1998}.
The ISAW theta point has been clearly identified as a
tricritical point in the language of $O(n)$ models, and all of
its properties are quite well-established, with good agreement
between theory~\cite{DuplantierSaleur1987} and
simulations~\cite{CaraccioloGherardiPapinuttoPelissetto2011}
for the values of critical exponents. Conversely, a number of
contradictory results have emerged for the ISAT and VISAW theta
points, in particular some striking discrepancies among Monte
Carlo
simulations~\cite{MeirovitchLim1988,OwczarekPrellberg2007_PHYSA,BediniOwczarekPrellberg2013_PRE,BediniOwczarekPrellberg2013_JPA2},
numerical transfer-matrix
methods~\cite{BloteNienhuis1989,FosterPinettes2003,Foster2009,Foster2011,FosterPinettes2012},
field-theoretic
arguments~\cite{JacobsenReadSaleur2003,NahumSernaSomozaOrtuno2013,VernierJacobsenSaleur2015},
and exact results~\cite{WarnaarBatchelorNienhuis1992}.

In this article, we study a generalized polymer model that
incorporates all the aforementioned ones, in the framework of
Husimi~\cite{Pretti2002,SerraStilckCavalcantiMachado2004,ZaraPretti2007,OliveiraStilckSerra2008,NetoStilck2013}
and Bethe
lattices~\cite{LiseMaritanPelizzola1998,BuzanoPretti2002,Pretti2006,SerraStilck2007,NetoStilck2008}.
The free parameters characterizing such lattices (building
blocks, coordination numbers) are chosen, according to
experience, in order to obtain the best possible approximation
to a regular 2d square lattice model. The motivation for this
work stems from the difficulty of extracting, from the existing
literature, a unified view of the phase diagram in the presence
of the different topological constraints and competing collapse
interactions described above. Indeed, considerable work has
been done in this direction in particular by Bedini, Owczarek,
and Prellberg (making use of refined Monte Carlo
simulations)~\cite{BediniOwczarekPrellberg2013_PRE,BediniOwczarekPrellberg2013_JPA1,BediniOwczarekPrellberg2014_JPA}
and by Foster and Pinettes (mainly by means of numerical
transfer-matrix
techniques)~\cite{FosterPinettes2003,Foster2009,Foster2011,FosterPinettes2012},
but several unresolved issues and contradictory results still
remains. Just to give a couple of examples, the mentioned
authors have investigated an \emph{asymmetric} ISAT (denoted as
AISAT) on the square lattice, where doubly-visited sites with
or without crossing are assigned a different attractive
interaction. Now, in the special case of crossing interactions
only, the transfer-matrix results~\cite{Foster2011} suggest the
onset of a first-order collapse, while the
simulations~\cite{BediniOwczarekPrellberg2013_PRE} seem to
predict a revival of the ISAW theta state, i.e., a weak
continuous transition, with no evidence of a maximally-dense
phase. Moreover, whilst an exact
solution~\cite{WarnaarBatchelorNienhuis1992} of the VISAW model
predicts a correlation-length exponent ${\nu = 12/23}$, Monte
Carlo simulations~\cite{BediniOwczarekPrellberg2013_JPA2}
suggest the ISAW value ${\nu = 4/7}$, and transfer-matrix
methods~\cite{BloteNienhuis1989,FosterPinettes2012} are
incompatible with both results.

In this context, a clear advantage of the approach developed in
the current article is the possibility of a sharp determination
of the phase diagram (with extremely high numerical precision
in the case of the Husimi lattice, even analytical in the case
of the Bethe lattice). For instance, we have no ambiguity on
the order of the transitions, which is usually not the case
with methods affected by higher numerical uncertainties. On the
other hand, a primary drawback of the matter is that the
resulting phase diagram, indeed exact on such
infinite-dimensional treelike lattices, is not guaranteed to
correspond to the actual phase diagram of the 2d model.
Concerning this point, we have nonetheless to say that Husimi
and Bethe lattice models often turn out to exhibit a remarkably
good qualitative agreement with finite-dimension results, when
the latter are well-established by other methods. This has been
observed in different polymer models, such as the semiflexible
ISAW~\cite{Pretti2002,LiseMaritanPelizzola1998,BastollaGrassberger1997}
or the
bond-ISAW~\cite{BuzanoPretti2002,FosterSeno2001,Foster2007},
and holds true, as we shall see, at least in a limiting case of
the model under
investigation~\cite{BediniOwczarekPrellberg2013_JPA1,BediniOwczarekPrellberg2014_JPA}.
Indeed, we shall see that the Bethe lattice model yields a
slightly less convincing behavior for the general case, but we
have found it interesting (and therefore we have included it in
the article), mostly by virtue of the opportunity of a fully
analytical solution.

A more substantial difficulty of our approach is that, even
though the Husimi or Bethe solutions take into account certain
local correlations, they still have a mean-field nature, so
that \emph{in principle} they cannot predict critical
exponents. As a consequence, the universal points are to be
naively identified on the basis of physical intuition, and/or
according to their role in the phase diagram. In conclusion, we
cannot expect that our results provide definitive answers to
the open problems of the 2d case, in particular to the issue of
universality classes, but we believe they might be nonetheless
of some use, mainly as a coherent set of hypotheses, yet to be
tested by more specific methods.

The paper is organized as follows. In Sec.~\ref{sec:model} we
give a precise definition of the model that we are going to
investigate. In Sec.~\ref{sec:husimi} we briefly present the
Husimi lattice solution, whose details are reported separately
in Appendix~\ref{app:rec}. Sec.~\ref{sec:results} contains all
the results concerning the Husimi lattice model, whereas
Sec.~\ref{sec:conclusions} is devoted to a discussion and some
concluding remarks. The Bethe lattice solution is reported in
full detail in Appendix~\ref{app:bethe}, along with a
comparison with the Husimi lattice one.


\section{The model}
\label{sec:model}

Let us define the model on the regular two-dimensional square
lattice, as in the original (partial) versions. The definition
for the Husimi or Bethe lattices follows in a straightforward
way. The polymer is represented as a \emph{self-avoiding
trail}, that is, a walk such that lattice bonds can be visited
only once, whereas lattice sites can be visited more than once
(at most twice on the square lattice). Doubly-visited sites are
assigned a Boltzmann weight $\tau_\mathrm{x}$ or
$\tau_\mathrm{c}$, depending on whether the walk
self-intersects or not, respectively. Moreover, a weight
$\omega$ is assigned to every pair of nearest-neighbor sites,
that are visited (once) by nonconsecutive monomers. The latter
type of statistical weight is the usual one for the ISAW. All
kinds of weights are summarized in Fig.~\ref{fig:model}. For
simplicity, in the following we shall denote the configurations
weighted by $\omega$ as \emph{contacts}, whereas those weighted
by $\tau_\mathrm{c}$ and $\tau_\mathrm{x}$ will be denoted as
\emph{collisions} and \emph{crossings}.

\begin{figure}
  \setlength{\unitlength}{0.8mm}
  \begin{picture}(90,90)(-29,-29)
    \linethickness{0.2\unitlength}
    \put(-28,-16){\line(1,0){88}}
    \put(-28,  0){\line(1,0){88}}
    \put(-28, 16){\line(1,0){88}}
    \put(-28, 32){\line(1,0){88}}
    \put(-28, 48){\line(1,0){88}}
    \put(-16,-28){\line(0,1){88}}
    \put(  0,-28){\line(0,1){88}}
    \put( 16,-28){\line(0,1){88}}
    \put( 32,-28){\line(0,1){88}}
    \put( 48,-28){\line(0,1){88}}
    \linethickness{0.8\unitlength}
    \put(-28,16){\line(1,0){26}}
    \qbezier(-2,16)(0,16)(0,14)
    \put(0,14){\line(0,-1){28}}
    \qbezier(0,-14)(0,-16)(2,-16)
    \put(2,-16){\line(1,0){12}}
    \qbezier(14,-16)(16,-16)(16,-14)
    \put(16,-14){\line(0,1){60}}
    \qbezier(16,46)(16,48)(18,48)
    \put(18,48){\line(1,0){28}}
    \qbezier(46,48)(48,48)(48,46)
    \put(48,46){\line(0,-1){28}}
    \qbezier(48,18)(48,16)(46,16)
    \put(46,16){\line(-1,0){44}}
    \qbezier(2,16)(0,16)(0,18)
    \put(0,18){\line(0,1){42}}
    \put(-1,18){\makebox[0pt][r]{$\Large \tau_\mathrm{c}$}}
    \put(18,18){\makebox[0pt][s]{$\Large \tau_\mathrm{x}$}}
    \put(8, 1){\makebox[0pt]{$\Large \omega$}}
    \put(8,33){\makebox[0pt]{$\Large \omega$}}
    \put(8,49){\makebox[0pt]{$\Large \omega$}}
  \end{picture}
  \caption
  {
    An example of polymer configuration on the 2d square lattice,
    along with the statistical weights associated to the different interactions:
    $\omega$ is associated to nearest-neighbor sites visited (once) by nonconsecutive
    monomers; $\tau_\mathrm{c}$ and $\tau_\mathrm{x}$ are associated
    to doubly-visited sites with, respectively, colliding or crossing configurations.
  }
  \label{fig:model}
\end{figure}

Let us briefly explain how this model incorporates different
polymer models, that have been previously studied in the
literature. First of all, when ${\tau_\mathrm{x} =
\tau_\mathrm{c} \equiv \tau}$ (i.e., collisions and crossings
are equally weighted), we obtain the Wu-Bradley
model~\cite{WuBradley1990}, which has recently been the subject
of an accurate numerical investigation by Bedini and
coworkers~\cite{BediniOwczarekPrellberg2014_JPA}. In turn, the
latter model contains the ordinary ISAT model for ${\omega =
1}$ (i.e., without contact interactions), the ordinary ISAW
model for ${\tau = 0}$ (i.e., forbidding doubly-visited sites),
and also the so-called \emph{interacting nearest-neighbor
self-avoiding trail} (INNSAT)
model~\cite{BediniOwczarekPrellberg2013_JPA1} for ${\tau = 1}$
(i.e., a self-avoiding trail with contact interactions only).
The latter model has been proposed in the cited article, in
order to investigate the stability of the ISAW theta state with
respect to a change in the self-avoidance contraints.
Furthermore, for ${\omega = 1}$ and ${\tau_\mathrm{x} \neq
\tau_\mathrm{c}}$, we obtain the previously mentioned AISAT
model~\cite{Foster2011,BediniOwczarekPrellberg2013_PRE},
studied in order to test the stability of the ISAT theta state
with respect to perturbations in the attractive interaction.
The latter model contains in turn the VISAW model, in the
limiting case ${\tau_\mathrm{x} = 0}$ (crossings forbidden).

We shall consider the model in a grand-canonical description,
with a fugacity parameter $\zeta$ associated to each polymer
segment, that is, to each visited bond. The partition function
can thus be written as
\begin{equation}
  \Xi = \sum_\mathrm{SAT}
  \omega^{\mathcal{N}_\mathrm{n}}
  \tau_\mathrm{c}^{\mathcal{N}_\mathrm{c}}
  \tau_\mathrm{x}^{\mathcal{N}_\mathrm{x}}
  \zeta^{\mathcal{N}_\mathrm{s}}
\end{equation}
where $\mathcal{N}_\mathrm{s}$ denotes the number of visited
bonds, and $\mathcal{N}_\mathrm{n,c,x}$ denote respectively the
number of contacts, collisions, and crossings. The sum is
understood to run over all configurations compatible with the
self-avoiding trail constraint. Note that, since we are
interested in the properties of infinitely long polymers, this
set of configurations does not include loops whose length
remains finite in the thermodynamic limit.


\section{The Husimi lattice solution}
\label{sec:husimi}

\begin{figure}
  \setlength{\unitlength}{0.8mm}
  \begin{picture}(90,90)(-29,-29)
    \linethickness{0.2\unitlength}
    \put(-28,  0){\line(1,0){88}}
    \put(  0,-28){\line(0,1){88}}
    \put(-28, 32){\line(1,0){88}}
    \put( 32,-28){\line(0,1){88}}
    \put(-28,-16){\line(1,0){40}}
    \put(-16,-28){\line(0,1){40}}
    \put(-28, 48){\line(1,0){40}}
    \put(-16, 20){\line(0,1){40}}
    \put( 20,-16){\line(1,0){40}}
    \put( 48,-28){\line(0,1){40}}
    \put( 20, 48){\line(1,0){40}}
    \put( 48, 20){\line(0,1){40}}
    \put( 20, 56){\line(1,0){16}}
    \put( 24, 44){\line(0,1){16}}
    \put( 44, 56){\line(1,0){16}}
    \put( 56, 44){\line(0,1){16}}
    \put( 44, 24){\line(1,0){16}}
    \put( 56, 20){\line(0,1){16}}
    \put(-28, 24){\line(1,0){16}}
    \put(-24, 20){\line(0,1){16}}
    \put(- 4, 56){\line(1,0){16}}
    \put(  8, 44){\line(0,1){16}}
    \put(-28, 56){\line(1,0){16}}
    \put(-24, 44){\line(0,1){16}}
    \put(-28,-24){\line(1,0){16}}
    \put(-24,-28){\line(0,1){16}}
    \put(- 4,-24){\line(1,0){16}}
    \put(  8,-28){\line(0,1){16}}
    \put(-28,  8){\line(1,0){16}}
    \put(-24,- 4){\line(0,1){16}}
    \put( 20,-24){\line(1,0){16}}
    \put( 24,-28){\line(0,1){16}}
    \put( 44,-24){\line(1,0){16}}
    \put( 56,-28){\line(0,1){16}}
    \put( 44,  8){\line(1,0){16}}
    \put( 56,- 4){\line(0,1){16}}
  \end{picture}
  \caption
  {
    Sketch of a Husimi lattice made up of square clusters.
  }
  \label{fig:husimilattice}
\end{figure}
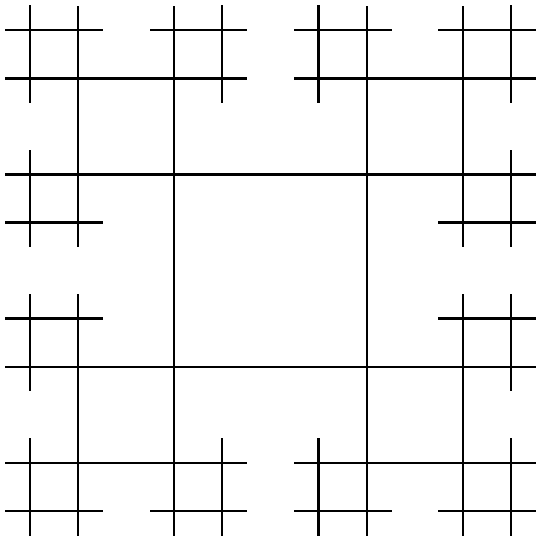

A \emph{Husimi tree} is a self-similar lattice like the one
depicted in Fig.~\ref{fig:husimilattice}, where it is
understood that the size of the tree is arbitrarily increased
(toward a thermodynamic limit) by a self-replicating growth
procedure. In a wide part of the literature, a \emph{Husimi
lattice} is defined as the ``inner region'' of a corresponding
infinite Husimi tree~\cite{LavisBell1999}, meaning a region
where a homogeneity condition holds for thermal averages of
local observables (for instance, the site magnetization for the
ferromagnetic Ising model). Such thermal averages can thus be
determined by the fixed points of relatively simple
self-consistency equations. Unfortunately, due to the fact that
in a Husimi tree the majority of sites is located on the
boundary, it turns out that in certain cases boundary
conditions heavily affect the properties of the inner region as
well, which is not a good fact to the purpose of approximating
the thermodynamic behavior of a regular lattice model. For a
discussion of these issues, see for instance
Refs.~\onlinecite{Pretti2003,Ostilli2012}. The Husimi lattice
is therefore better defined~\cite{DorogovtsevGoltsevMendes2008}
as an ensemble of random-regular hyper-graphs, that are,
roughly speaking, random graphs made up of a unique type of
hyper-edge (for instance, a square cluster as in
Fig.~\ref{fig:husimilattice}), with a fixed coordination
number. This kind of systems have no boundary, so that the
problem of boundary conditions is avoided, and are
\emph{locally treelike}, since the length of closed paths
diverges as the logarithm of the number of sites in the
thermodynamic limit (excluding of course short paths that are
closed within single clusters). The latter fact means that a
Husimi lattice still looks (locally) like
Fig.~\ref{fig:husimilattice} and, more importantly, that its
thermodynamic properties can still be derived in terms of
simple self-consistency equations. Of course, our choice to use
square clusters as building blocks, and a coordination number
$2$ (in this context, we define the coordination number as the
number of building blocks attached to each given site) is
motivated by the purpose of approximating the model on the
ordinary 2d square lattice. The self-consistency equations
(a.k.a. \emph{recursion equations}) can be worked out in
different ways. A possible way, which we believe likely to be
the simplest one, is to derive them as stationarity conditions
for a suitable variational free energy density.

Let us consider the grand-canonical free-energy density per
site (in $k_\mathrm{B}T$ units)
\begin{equation}
  \psi = - \lim_{N \to \infty} \frac{1}{N} \ln \Xi
  \, ,
\end{equation}
where $N$ denotes the number of sites. For the Husimi lattice
defined above, this free-energy density can be written as
\begin{equation}
  \psi = - \frac{1}{2} \ln q + \ln z
  \, ,
  \label{eq:free-energy_density}
\end{equation}
where $q$ is the \emph{cluster partition function}, $z$ is the
\emph{single-site partition function}, and $1/2$ is the ratio
between the number of square clusters and the number of sites
present in the system. Roughly speaking, the single-site term
may be regarded as a correction over the cluster term, due to
the overlap between clusters.

The cluster partition function $q$ is
the partition function of a small subsystem
made up of four sites on a square cluster,
interacting with one another and with effective fields
that represent the remainder of the system.
It will therefore take the following form
\begin{equation}
  q = \sum_{i,j,k,l} \chi_{i,j,k,l} w_{i} w_{j} w_{k} w_{l}
  \, ,
  \label{eq:cluster_partition_function}
\end{equation}
where $i,j,k,l$ are labels for the polymer configurations on
each site, $\chi_{i,j,k,l}$ is the statistical weight of
interactions inside the square cluster (including topological
constraints), and $w_{i}$ are the effective fields. Note that
these fields are sometimes called \emph{cavity fields}, since
they represent the probabilities of site configurations for a
system in which a cluster interaction is removed (that is,
replaced by a cavity). Moreover, ${- \ln q}$ can be regarded as
the \emph{free-energy shift} between the system with a cavity
and the unperturbed system. For clarity, let us note that, in
most papers dealing with polymer models on Husimi or Bethe
lattices, the cavity fields are called \emph{partial partition
functions}. In equation~\eref{eq:cluster_partition_function}
the treelike nature of the system is reflected in the fact
that, in the absence of the cluster interaction, the joint
probability distribution factorizes, meaning that the site
configurations are statistically independent. Indeed, in a real
tree-graph the removal of a cluster splits the system into
noninteracting subsystems (usually called \emph{branches}).

The single-site partition function $z$ is a similar quantity
for a single site, interacting with two cavity fields
associated to the two ``branches'' attached to the given site.
It can be written in the following form
\begin{equation}
  z = \sum_{i,j} \chi_{i,j} w_{i} w_{j}
  \, .
  \label{eq:site_partition_function}
\end{equation}
Note that we have two configuration labels $i,j$ even though
for a single site, because it is convenient to define the local
polymer configurations with respect to two different reference
frames, integral with each branch. As a consequence, the same
configuration may be identified by different labels with
respect to different branches. Moreover, the statistical weight
$\chi_{i,j}$ also has to ensure consistency between the two
labels. This issue should get clearer in
Appendix~\ref{app:rec}, where we report the explicit expression
of~$z$.

All the information needed to solve the model is contained in
equations \eref{eq:free-energy_density},
\eref{eq:cluster_partition_function},
and~\eref{eq:site_partition_function}, provided explicit
expressions of the statistical weights $\chi_{i,j,k,l}$ and
$\chi_{i,j}$ (and therefore of $q$ and $z$) are derived. As
previously mentioned, equilibrium states are determined by the
stationarity conditions for the free energy, namely,
\begin{equation}
  \frac{\partial \psi}{\partial w_{i}} = 0
  \, ,
\end{equation}
which can be easily written in a self-consistent form
\begin{equation}
  w_{i} \propto f_{i}(w_{0},w_{1},w_{2},w_{3})
  \label{eq:rec}
  \, .
\end{equation}
Explicit expressions for the functions $f_i$ are also reported
in Appendix~\ref{app:rec}, along with a proof that 4 cavity
fields (denoted as ${w_{0},\dots,w_{3}}$) suffice to represent
all the relevant configurations. The recursion
equations~\eref{eq:rec} can be solved numerically by a simple
iterative algorithm. Note that a proportionality constant
remains undetermined, because the free-energy density $\psi$ is
invariant under multiplication of each cavity field by a
constant, as one can easily verify. As usual, we fix the
constant so as to satisfy, at each iteration, a normalization
condition, namely
\begin{equation}
  \sum_{i=0}^{3} w_{i} = 1
  \, .
  \label{eq:norm}
\end{equation}
For the benefit of readers who are not familiar with Husimi
lattice models, let us briefly note that the self-consistent
form is not only an efficient way to solve numerically a set of
nonlinear simultaneous equations, but it can be interpreted as
a self-similarity condition, namely, the equality between the
cavity fields associated to a given branch and those of its
sub-branches. Indeed, most studies directly derive the
recursion equations via self-similarity.

All equilibrium properties of the system can now be derived
from the knowledge of the cavity field values satisfying
equations~\eref{eq:rec}, along with the expression of the
free-energy density, equations \eref{eq:free-energy_density},
\eref{eq:cluster_partition_function},
and~\eref{eq:site_partition_function}. The average number of
segments per site, which in the following we shall briefly
refer to as {\em segment density}, or simply \emph{density},
can be evaluated as
\begin{equation}
  \rho = - \frac{\partial \psi}{\partial \ln \zeta}
  \, .
  \label{eq:rho}
\end{equation}
Note that, in the above derivative, the dependence on~$\zeta$
of the cavity fields can be neglected, because we are
interested in equilibrium points, and the derivatives of $\psi$
with respect to the cavity fields vanish at such points. The
density~$\rho$ is the main order parameter for our system.
Other meaningful observables are the average number of contacts
per site (\emph{contact density})
\begin{equation}
  \eta = - \frac{\partial \psi}{\partial \ln \omega}
\end{equation}
and the average number of collisions or crossings per site
(respectively, \emph{collision} or \emph{crossing density})
\begin{equation}
  \xi_\mathrm{c,x}
  = - \frac{\partial \psi}{\partial \ln \tau_\mathrm{c,x}}
  \, .
\end{equation}
Explicit expressions for $\xi_\mathrm{c}$, $\xi_\mathrm{x}$, and $\eta$
can be derived straightforwardly.

In the presence of multiple solutions (i.e., fixed points) of
the recursion equations, revealing the occurrence of
coexistence phenomena, the free-energy values allow one to
discriminate the thermodynamically stable phase, and therefore
to determine first-order transitions. Conversely, second-order
transitions can be better detected by analyzing the {\em
stability} of the solutions. The latter is a rather technical
issue, which we discuss in Appendix~\ref{app:rec}.


\section{Results}
\label{sec:results}

In the framework of the grand-canonical formulation, the phase
diagram can be described as a function of the Boltzmann weights
of the elementary interactions, also denoted as
\emph{activities} (in our case $\tau_\mathrm{c}$,
$\tau_\mathrm{x}$, and $\omega$), and of the \emph{fugacity}
variable $\zeta$, which controls the segment density. For given
activity values, we expect a phase transition to occur at a
certain fugacity
$\zeta_{0}(\tau_\mathrm{c},\tau_\mathrm{x},\omega)$, where the
density changes from ${\rho = 0}$ for ${\zeta < \zeta_{0}}$ to
${\rho > 0}$ for ${\zeta > \zeta_{0}}$. The transition
manifold, defined by the function $\zeta_{0}$, can be
identified as the canonical thermodynamic limit for a single
polymer chain, and in particular ${\ln \zeta_{0}}$ represents
the corresponding Helmholtz free-energy density per segment (in
${k_\mathrm{B} T}$ units). In the limit ${\zeta \to
\zeta_{0}^{+}}$, the properties of the dense phase are expected
to approach those of a single chain (we shall denote it as the
``single-chain'' limit), so that, in particular, the segment
density~$\rho$ can be viewed as a measure of the chain
compactness. Therefore, a second-order transition represents a
swollen state, whereas a first-order transition represents a
collapsed state. For the well-known ISAW model
(${\tau_\mathrm{x} = \tau_\mathrm{c} = 0}$ in our scheme),
increasing the contact activity~$\omega$ (which usually means
lowering the temperature) drives the system from the former to
the latter regime, giving rise to a continuous transition
(theta collapse).

Solving the recursion equations~\eref{eq:rec}, we find, besides
the zero-density phase (also denoted as \emph{0~phase} in the
following), two different dense phases: an ``ordinary'' one
(\emph{1~phase}), and a ``maximally dense'' one
(\emph{2~phase}). The 2~phase is characterized by very large
values of the segment density $\rho$ and of the
crossing-collision density ${\xi \equiv \xi_\mathrm{c} +
\xi_\mathrm{x}}$ (quite close to the respective upper bounds
${\rho = 2}$ and ${\xi = 1}$), and by small values of the
contact density $\eta$ (quite close to $0$). Conversely, the
1~phase is characterized by smaller values of $\rho$ and $\xi$,
and larger values of $\eta$. The segment density $\rho$ is the
natural order parameter for both 0-1 and 0-2 phase transitions,
because the 0~phase is characterized by ${\rho = 0}$, whereas
${\rho > 0}$ in the 1 and 2 phases. The 1-2 transition cannot
be defined rigorously in terms of densities, but a suitable
order parameter is instead the $w_1$ cavity field, which turns
out to be strictly positive in the 1~phase and zero in the
2~phase.

\subsection{Wu-Bradley model}

Let us first analyze the case of a symmetric crossing-collision
interaction (${\tau_\mathrm{x} = \tau_\mathrm{c} \equiv
\tau}$), which has been previously denoted as the Wu-Bradley
model. We shall present a sequence of projections of the
grand-canonical phase diagram in the plane $\zeta$~vs~$\tau$,
for different fixed values of the contact activity, namely,
${\omega = 1.2, 0.5, 1}$. Note that ${\omega
> 1}$ or ${\omega < 1}$ means that contacts are energetically
favored or disfavored, respectively, whereas the case ${\omega
= 1}$ corresponds to the ISAT model, with no contact
interaction.

\begin{figure}
  \resizebox{85mm}{!}{\includegraphics*{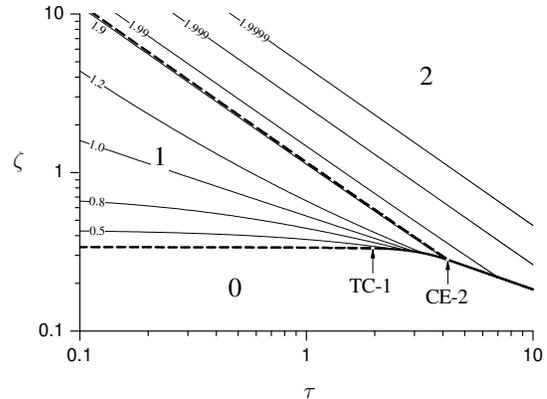}}
  \caption
  {
    Grand-canonical phase diagram ($\zeta$~vs~$\tau$)
    for ${\omega = 1.2}$.
    The zero-density phase, the ordinary dense phase,
    and the maximally-dense phase are denoted respectively by 0, 1, 2.
    Solid lines denote first order transitions;
    dashed lines denote second order transitions.
    Thin solid lines are contour lines of the segment density per site
    (numerals denoting density values).
    The point at which the 0-1 transition changes
    from second to first order is tagged as TC-1.
    The critical end-point, terminating the (second-order)
    1-2 transition line, is tagged as CE-2.
  }
  \label{fig:omega=1-2}
\end{figure}

The phase diagram for ${\omega = 1.2}$ is displayed in Fig.~\ref{fig:omega=1-2}.
The transition line between the 0 and 1 phases turns out to be
partially first and partially second order. The point
separating the two regimes represents a continuous collapse in
the single-chain limit, so that it can be viewed as the
analogous of the theta point of the ISAW model, which is, a
tricritical point in the language of $O(n)$ models. We shall
denote this point as TC-1 to avoid misunderstandings, because,
as previously mentioned, the theta point of the ISAW model
corresponds to a well-defined universality class, whereas
Husimi lattice models necessarily belong to a mean-field
universality class. In the dense region (${\rho > 0}$) a
second-order transition line separates the 1 and 2 phases. This
line joins to the transition line with the 0 phase at a
critical end-point, which we shall denote as CE-2. The latter
corresponds, in the single-chain limit, to a continuous
transition between two different regimes of the collapsed
state. The regime associated to the first-order portion of the
0-1 phase represents a moderately compact state, whose density
rapidly increases upon increasing $\tau$. On the other hand,
the regime associated to the 0-2 transition (which is all
first-order) represents a very compact state, with a large
majority of doubly-visited sites, whose density value is almost
saturated around ${\rho \lesssim 2}$. These observations can be
confirmed by inspection of the density contour lines, reported
in Fig.~\ref{fig:omega=1-2}.

\begin{figure}
  \resizebox{85mm}{!}{\includegraphics*{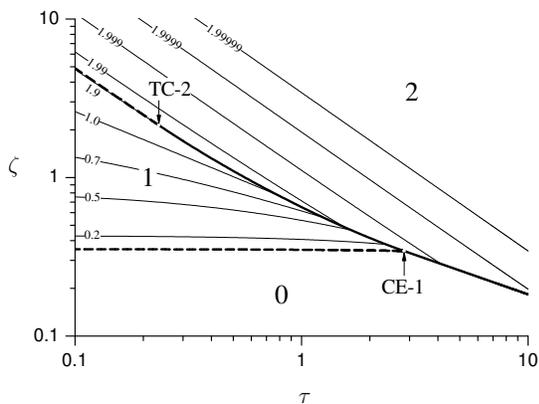}}
  \caption
  {
    Grand-canonical phase diagram ($\zeta$~vs~$\tau$)
    for ${\omega = 0.5}$.
    Lines and phase tags are as in the previous figure.
    The critical end-point, that terminates the (second-order)
    0-1 transition line, is tagged as CE-1.
    The point at which the 1-2 transition changes
    from second to first order is tagged as TC-2.
  }
  \label{fig:omega=0-5}
\end{figure}

The transition scenario changes considerably for ${\omega < 1}$.
Fig.~\ref{fig:omega=0-5} displays the $\zeta$ vs $\tau$ phase diagram
in the particular case ${\omega = 0.5}$.
The TC-1 point disappears, so that the 0-1 transition line is
now all second-order. Conversely, the 1-2 transition line turns
out to be partially second and partially first order. As a
consequence, the 0-1 transition line terminates in a critical
end-point, which we denote as CE-1. The latter also marks the
separation between the first-order portion of the 1-2
transition line and the (all first order) 0-2 transition line.
In the single-chain limit the CE-1 point represents a
discontinuous (first order) collapse transition to the very
compact state with almost saturated density.

\begin{figure}
  \resizebox{85mm}{!}{\includegraphics*{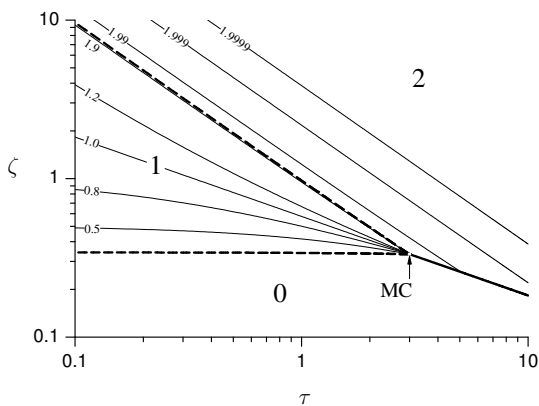}}
  \caption
  {
    Grand-canonical phase diagram ($\zeta$~vs~$\tau$)
    for ${\omega = 1}$ (ISAT model).
    Lines and phase tags are as in the previous figures.
    The multicritical point is tagged as MC.
  }
  \label{fig:omega=1-0_ISAT-model}
\end{figure}

Analyzing the evolution of the $\zeta$~vs~$\tau$ phase diagram
between the two regimes observed, from ${\omega = 1.2}$ to
${\omega = 0.5}$, we realize that all the relevant changes
occur precisely at ${\omega = 1}$, that is, in the case of the
pure ISAT model. The phase diagram for this case is reported in
Fig.~\ref{fig:omega=1-0_ISAT-model}.
We can observe that all four of the ``special points'' defined
above, namely, TC-1, TC-2, CE-1, and CE-2 degenerate into a
unique multicritical point (tagged as MC). In the single-chain
limit, this point still represents an abrupt collapse
transition to the saturated compact state, but it turns out to
exhibit rather peculiar features. First of all, it is located
precisely at ${\tau = 3}$ and ${\zeta = 1/3}$. These values,
which we can determine with very high numerical precision (of
the order of ten decimal places), coincide with those pointed
out (though with lower precision) by both
simulations~\cite{BediniOwczarekPrellberg2014_JPA} and
transfer-matrix techniques~\cite{Foster2009}. Furthermore, we
can observe that all the density contour lines in the 1~phase
converge toward the MC point. This means that precisely
\emph{at} the MC point there exists a continuum of solutions of
the recursion equations (i.e., a continuum of minima of the
free energy), all with the same free-energy value, but with
densities ranging from zero (the 0-phase value) up to the
2-phase value. In other words, we can state that the 0 phase
and the 2 phase remain distinct, but the free energy barrier
associated with the 0-2 (first-order) transition vanishes
precisely at the MC point. As far as the single-chain limit is
concerned, we might analogously say that the collapse
transition of the ISAT model is a discontinuous transition with
a zero free-energy barrier. Finally, it is noticeable that both
the location and the specific features of the MC point are
preserved even in the Bethe lattice solution. In this case all
the details reported above can be probed analytically, as
discussed in Appendix~\ref{app:bethe}. In particular, it turns
out that, in a Landau expansion of the variational free energy,
the derivatives of any order vanish precisely at the MC point.
This is just the mean-field representation of a multicritical
point of infinite order, which remarkably agrees with the
characterization of the ISAT theta point, recently given by
Nahum and coworkers.~\cite{NahumSernaSomozaOrtuno2013}

\begin{figure}
  \resizebox{85mm}{!}{\includegraphics*{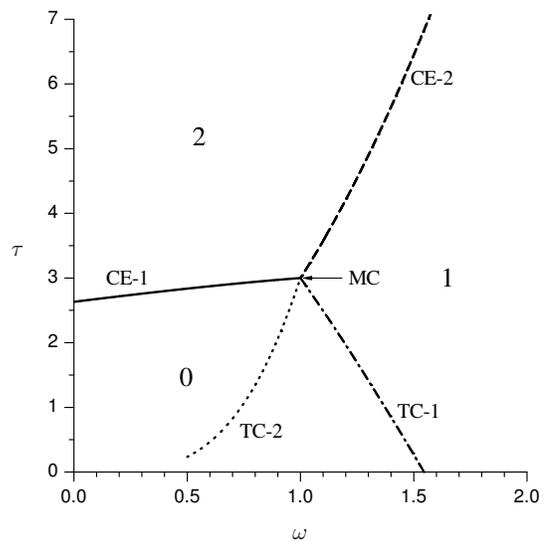}}
  \caption
  {
    Single-chain phase diagram ($\tau$~vs~$\omega$).
    Here 0, 1, and 2 denote respectively the swollen state,
    the moderately-compact and the highly-compact states.
    The solid line and the dashed line represent the evolution of
    the CE-1 and CE-2 points, respectively.
    The evolution of the TC-1 and TC-2 points is represented by
    the dash-dotted and dotted lines, respectively.
    Note that TC-2 points lie outside the 0-phase boundary,
    and do not represent single-chain transitions.
  }
  \label{fig:omegatau}
\end{figure}

To complete the picture of the phase diagram for the Wu-Bradley
model, we have systematically analyzed the evolution of the
``special points'' as a function of $\omega$. This analysis
leads to the graph reported in Fig.~\ref{fig:omegatau}, which
can also be regarded as a single-chain phase diagram.
Indeed, all the special points, except TC-2, lie on the ${\zeta
= \zeta_0}$ manifold (i.e., on the 0-phase boundary), and
represent certain conformational transitions for a single
polymer chain in the thermodynamic limit (note that all the
points except MC evolve into lines upon varying $\omega$). In
particular: TC-1 represents a continuous ISAW-like collapse
from a swollen state to a moderately compact state, CE-2
represents another continuous collapse from the moderately
compact to the highly compact or saturated state; CE-1
represents a direct discontinuous (first-order) collapse from
the swollen state to the highly compact state. Finally, MC can
be regarded as a limiting case of CE-1, in which the
free-energy barrier vanishes.

\subsection{Generalized model and AISAT model}

Let us now switch to the general case ${\tau_\mathrm{x} \neq
\tau_\mathrm{c}}$. Let us note that, in principle, a complete
description of the phase diagram would require the exploration
of the whole four-parameters space
($\zeta,\omega,\tau_\mathrm{c},\tau_\mathrm{x}$), which would
be a rather heavy task. Nevertheless, it turns out that the
description can be considerably simplified by introducing
suitable alternative parameters, namely, a particular
\emph{average crossing-collision activity}
\begin{equation}
  \tau  \equiv  \frac{2 \tau_\mathrm{c} + \tau_\mathrm{x}}{3}
  \, ,
  \label{eq:tau}
\end{equation}
and an \emph{asymmetry parameter}
\begin{equation}
  \Gamma  \equiv  \frac{\tau_\mathrm{x} - \tau_\mathrm{c}}{\tau_\mathrm{x} + \tau_\mathrm{c}}
  \, .
  \label{eq:gamma}
\end{equation}
The range of the latter turns out to be the interval $[-1,1]$,
where ${\Gamma = 0}$ corresponds to the symmetric case
${\tau_\mathrm{x} = \tau_\mathrm{c} = \tau}$ (Wu-Bradley
model), whereas ${\Gamma = +1, -1}$ correspond respectively to
the extreme asymmetric cases ${\tau_\mathrm{c} = 0}$ and
${\tau_\mathrm{x} = 0}$ (VISAW model).

\begin{figure}
  \resizebox{85mm}{!}{\includegraphics*{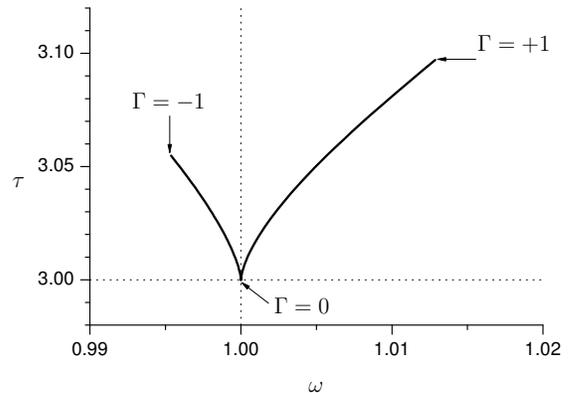}}
  \caption
  {
    Parametric plot of the position of the MC point
    in the $\tau$~vs~$\omega$ plane, as a function of $\Gamma$.
    Special $\Gamma$ values are indicated by arrows.
  }
  \label{fig:omegatau_gamma}
\end{figure}

The key-point is that, for fixed $\Gamma$ values, the
single-chain phase diagram in the $\tau$~vs~$\omega$ plane
remains topologically equivalent to that of the symmetric case
(Fig.~\ref{fig:omegatau}), with just a small displacement of
the MC-point location $(\omega_\mathrm{MC},\tau_\mathrm{MC})$
with respect to its ``original'' position $(1,3)$. A parametric
plot of $(\omega_\mathrm{MC},\tau_\mathrm{MC})$ as a function
of $\Gamma$ is reported in Fig.~\ref{fig:omegatau_gamma}.
We notice a peculiar cusp around the point $(1,3)$, which can
be characterized as follows. We verify numerically that for
small $\Gamma$, the MC-point coordinates behave as
\begin{eqnarray}
  \tau_\mathrm{MC} - 3 & \sim & \Gamma^2
  \, , \\
  \omega_\mathrm{MC} - 1 & \sim & \Gamma^3
  \, .
\end{eqnarray}
This obviously imply
\begin{eqnarray}
  \tau_\mathrm{MC} - 3 & \sim & \left| \omega_\mathrm{MC} - 1 \right|^{2/3}
  \, ,
\end{eqnarray}
which is indeed represented by a cusp in the $\tau$~vs~$\omega$
plane.

Apart from details, the plot of Fig.~\ref{fig:omegatau_gamma}
gives us some important information also about the behavior of
the AISAT model, which corresponds to the plane ${\omega = 1}$.
Keeping in mind the shape of the $\tau$~vs~$\omega$ phase
diagram (Fig.~\ref{fig:omegatau}), we can argue that, as soon
as ${\Gamma > 0}$ (i.e., ${\tau_\mathrm{c} <
\tau_\mathrm{x}}$), the ${\omega = 1}$ line will cross the CE-1
line, which means that the AISAT polymer undergoes a
first-order like collapse from the swollen state to the
highly-compact state. Conversely, when  ${\Gamma < 0}$ (i.e.,
${\tau_\mathrm{c} > \tau_\mathrm{x}}$), the ${\omega = 1}$ line
will cross both the TC-1 line and the CE-2 line, which means
that the AISAT polymer undergoes two different continuous
transitions, from the swollen state to the moderately-compact
state, and then to the highly-compact one. These two transition
scenarios are qualitatively equivalent to those previously
reported for the Wu-Bradley model, respectively for ${\omega >
1}$ and ${\omega < 1}$. Indeed, such an equivalence appears
rather intuitive, because a collision interaction is somehow
similar, on a shorter length scale, to a contact interaction
(even though the former necessarily implies polymer bending,
whereas the latter does not). As a consequence, a symmetric
ISAT model (${\tau_\mathrm{x} = \tau_\mathrm{c} = \tau}$) with
the addition of a favored nearest-neighbor contact interaction
(${\omega > 1}$) can be argued to be similar to an AISAT model
with collisions preferred to crossings (${\tau_\mathrm{c}
> \tau_\mathrm{x}}$). Analogously, a symmetric ISAT model with
the addition of a disfavored contact interaction (${\omega <
1}$) appears similar to an AISAT model with collisions
penalized with respect to crossing (${\tau_\mathrm{c} <
\tau_\mathrm{x}}$).

\begin{figure}
  \resizebox{85mm}{!}{\includegraphics*{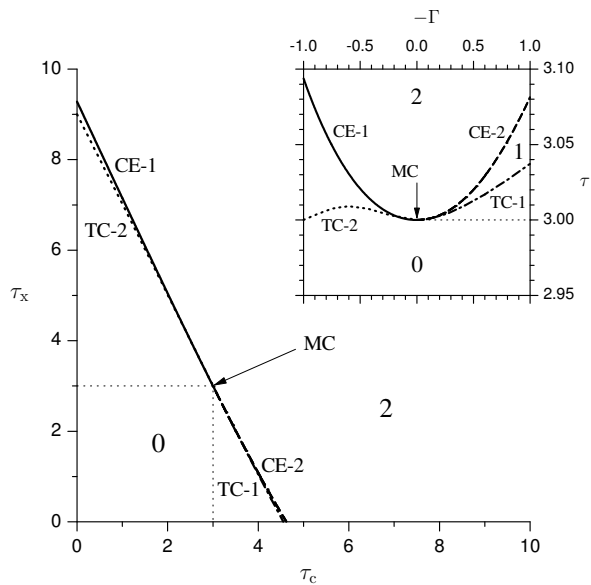}}
  \caption
  {
    Single-chain phase diagram for the AISAT model
    ($\tau_\mathrm{x}$ vs $\tau_\mathrm{c}$ in the main graph
    and $\tau$ vs $-\Gamma$ in the inset).
    Lines and tags are defined as in Fig.~\ref{fig:omegatau}.
  }
  \label{fig:tauctaux}
\end{figure}

\begin{figure}
  \resizebox{85mm}{!}{\includegraphics*{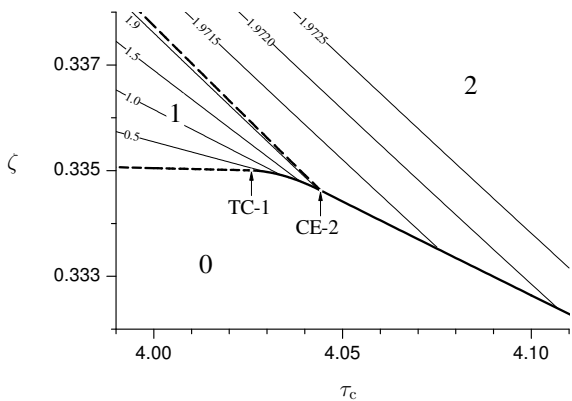}}
  \caption
  {
    Grand-canonical phase diagram ($\zeta$ vs $\tau_\mathrm{c}$)
    for ${\omega = 1}$ and ${\tau_\mathrm{x} = 1}$ (colliding model).
    Lines and tags are as in previous figures.
  }
  \label{fig:tau-x=1_collision-model}
\end{figure}

\begin{figure}
  \resizebox{85mm}{!}{\includegraphics*{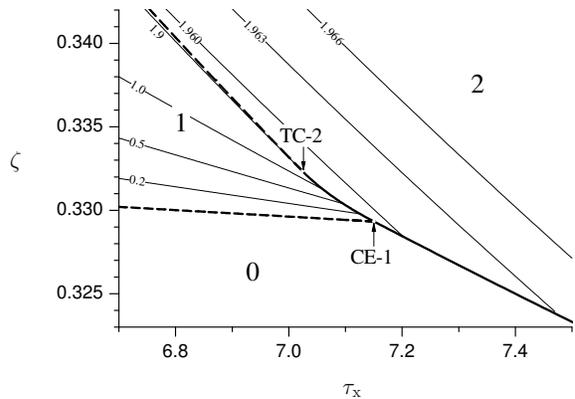}}
  \caption
  {
    Grand-canonical phase diagram ($\zeta$ vs $\tau_\mathrm{x}$)
    for ${\omega = 1}$ and ${\tau_\mathrm{c} = 1}$ (crossing model).
    Lines and tags as in previous figures.
  }
  \label{fig:tau-c=1_crossing-model}
\end{figure}

We have devoted special attention to the AISAT model, because
it has recently been the subject of different studies on the
ordinary 2d square
lattice~\cite{Foster2011,BediniOwczarekPrellberg2013_PRE}. In
Fig.~\ref{fig:tauctaux} we report the single-chain phase
diagram for the Husimi lattice case.
In the $\tau_\mathrm{x}$ vs $\tau_\mathrm{c}$ diagram, the
collapse transition looks like a single line passing through
the MC point (which occurs for ${\tau_\mathrm{x} =
\tau_\mathrm{c} = 3}$), separating the swollen state from the
highly-compact state. Nevertheless, the mapping on the
alternative parameters $\tau$ and $\Gamma$
(Fig.~\ref{fig:tauctaux} inset) reveals the fine structure of
this ``line''. In fact, for ${\tau_\mathrm{c} >
\tau_\mathrm{x}}$, the transition line splits into two
different lines of TC-1 and CE-2 points, whereas for
${\tau_\mathrm{c} < \tau_\mathrm{x}}$ it turns out to be a line
of CE-1 points, which is, a first-order collapse. The
similarity of the $\tau$~vs~$-\Gamma$ diagram for the AISAT
model with the $\tau$~vs~$\omega$ diagram of the Wu-Bradley
model is now evident. The most relevant difference is that, in
the latter case, the TC-1 and CE-2 lines become well separated
for large $\omega$ values, whereas, in the former case, they
remain very close to each other over the whole range of
admissible $\Gamma$ values. In Figs.
\ref{fig:tau-x=1_collision-model}
and~\ref{fig:tau-c=1_crossing-model} we report grand-canonical
phase diagrams for the two special cases ${\tau_\mathrm{x} =
1}$ and ${\tau_\mathrm{c} = 1}$, respectively.
This two cases have been respectively denoted as
\emph{colliding model} and \emph{crossing
model}~\cite{BediniOwczarekPrellberg2013_PRE}, since either
only collisions or only crossings are energetically favored,
even though \emph{both} collisions and crossings are always
allowed. We can fully appreciate the analogy with the ${\omega
> 1}$ and ${\omega < 1}$ cases of the Wu-Bradley model (Figs.
\ref{fig:omega=1-2} and~\ref{fig:omega=0-5}, respectively),
even though at a much smaller energy scale.


\section{Discussion and conclusions}
\label{sec:conclusions}

Let us now discuss our results, with special attention to
comparing them with those appeared in the literature for
corresponding two-dimensional models.

As far as the Wu-Bradley model (${\tau_\mathrm{x} =
\tau_\mathrm{c}}$) is concerned, we have obtained a nice
agreement with the (single-chain) phase diagram proposed by
Bedini and coworkers~\cite{BediniOwczarekPrellberg2014_JPA}.
The 0-2 transition (CE-1 line in Fig.~\ref{fig:omegatau}) is
conjectured to be first order, in agreement with our
prediction. Moreover, the 0-1 transition is conjectured to be
in the ISAW-theta universality class, and, even though in this
case we cannot give a corresponding prediction for critical
exponents, we find it significant that our 0-1 transition line
is indeed made up of tricritical (TC-1) points. Note also that
for ${\tau = 0}$ (ISAW model) the TC-1 point occurs at ${\omega
\approx 1.55}$, to be compared with the theta point of the 2d
square lattice model, occurring at ${\omega \approx
1.95}$~\cite{CaraccioloGherardiPapinuttoPelissetto2011}.
Finally, the 1-2 transition is not characterized precisely in
the cited article, due to difficulties in simulating dense
polymers, but there is reasonable evidence that it is a
continuous transition, which is also in agreement with our
prediction. On top of this, Bedini and coworkers identify the
merging point of the above transitions with the ISAT collapse,
characterized by specific properties (putatively, an
infinite-order multicritical
point~\cite{NahumSernaSomozaOrtuno2013}), which disappear as
soon as an arbitrarily small contact interaction is introduced.
There is indeed a striking analogy with the MC~point of our
Husimi-lattice model. We also find it noticeable that the
Husimi-lattice solution reproduces some fine details of the 2d
phase diagram, for instance, the slight positive slope of the
CE-1 line and the precise location of the MC point.

As far as the AISAT model (${\omega = 1}$) is concerned, the
situation is more involved. For the case of preferred crossings
(${\tau_\mathrm{c} < \tau_\mathrm{x}}$), our prediction, that
the collapse transition becomes first order, agrees with
Foster's transfer-matrix results~\cite{Foster2011}. In the
cited reference, it is not possible to estimate the precise
point at which the change of regime takes place, but the
results seem to be compatible with our scenario. Unfortunately,
the first-order transition is not confirmed by Bedini and
coworkers' simulations~\cite{BediniOwczarekPrellberg2013_PRE},
which conversely predict, for the \emph{crossing model}
(${\tau_\mathrm{c} = 1}$), an ISAW-like transition, with a
moderately compact collapsed state. Let us note that, probably
with an eye to Foster's findings, the latter authors report
having observed some evidence of a multiply-peaked probability
distribution, which seems nonetheless to become unimodal at
large lengths. Our scenario might be compatible with such not
fully definite results, because the close proximity of the TC-2
point (where the 1-2 transition becomes critical) to the CE-1
point (see Fig.~\ref{fig:tau-c=1_crossing-model}) implies a
very weak free-energy barrier, which might in fact give rise to
a broad probability distribution, rather than two sharp peaks.
Note indeed in Fig.~\ref{fig:omega=0-5} the much larger
distance between the TC-2 and CE-1 points for the Wu-Bradley
model with ${\omega = 0.5}$, where the first-order transition
is clearly pointed out by the
simulation~\cite{BediniOwczarekPrellberg2014_JPA}.

In the opposite case of preferred collisions (${\tau_\mathrm{c}
> \tau_\mathrm{x}}$), a comparison with 2d findings presents
even more difficulties. Our results predict a two-stage
collapse, with a TC-1 point followed by a CE-2 point (see
Fig.~\ref{fig:tau-x=1_collision-model}). As mentioned in the
text, both these points correspond to continuous transitions,
the former (which we have conjectured to be ISAW-like) toward a
moderately-compact state, and the latter toward a
highly-compact one. Again, due to the close proximity of the
two points in the parameters space, we believe that such a
scenario is not incompatible with Bedini and coworkers'
simulations~\cite{BediniOwczarekPrellberg2013_PRE}, predicting
a highly-dense collapsed phase for the \emph{colliding model}
(${\tau_\mathrm{x} = 1}$). On the other hand, Foster's
transfer-matrix results~\cite{Foster2011} seem to suggest that
the ISAT critical properties extend to the whole range of
preferred collisions (or penalized crossings) ${\tau_\mathrm{c}
\geq \tau_\mathrm{x} > 0}$, but the validity of these results
(for the pure ISAT case itself~\cite{Foster2009}) has recently
been questioned, by means of field-theoretic
arguments~\cite{NahumSernaSomozaOrtuno2013}, pointing out the
presence of very strong logarithmic corrections.

The limiting case ${\tau_\mathrm{x} = 0}$ (VISAW model) is
worth a separate discussion, because in this case an exact
solution~\cite{WarnaarBatchelorNienhuis1992} of the
corresponding $O(n)$ model (due to Bl\"{o}te and
Nienhuis~\cite{BloteNienhuis1989}) points out a specific
universality class, different from that of the ISAW. The
critical exponent values predicted by the exact solution turn
out to be quite elusive, since transfer-matrix
approaches~\cite{BloteNienhuis1989,FosterPinettes2012} give
incompatible results, whereas Monte Carlo
simulations~\cite{BediniOwczarekPrellberg2013_JPA2} are almost
compatible (at least for the $\nu$ exponent) with the ISAW
universality class itself. A careful investigation of the
Bl\"{o}te-Nienhuis theta state, recently performed by Vernier,
Jacobsen, and Saleur~\cite{VernierJacobsenSaleur2015}, suggests
that so difficult numerical estimates might arise from the
peculiar nature of the associated conformal field theory,
characterized by a continuous spectrum of critical exponents.
In order to reconcile our phase diagram for the VISAW case (in
fact practically equivalent to that of
Fig.~\ref{fig:tau-x=1_collision-model}) with the exact results,
we should postulate that the actual collapse transition is
indeed the TC-1 point, so that the Bl\"{o}te-Nienhuis theta
state should be associated to a different object, which we
might tentatively identify with the CE-2 point. Of course, on
the sole basis of our arguments, this possibility is nothing
more than speculation, but we find it quite intriguing, because
it seems not to be inconsistent with the most up-to-date
results~\cite{VernierJacobsenSaleur2015}, still inconclusive
about the role of the different integrable points in the phase
diagram. On the other hand, a slight evidence supporting our
interpretation is provided by the fact that the
Bl\"{o}te-Nienhuis point is characterized by a nonzero
density~\cite{FosterPinettes2003,VernierJacobsenSaleur2015},
and that, approached upon increasing fugacity, it corresponds
to a discontinuous transition~\cite{VernierJacobsenSaleur2015}.
Both features qualitatively agree with our predictions for the
CE-2 point.

Let us finally spend a few words about the analytical technique
that we have used in this article. Even though the solution of
Husimi and Bethe lattice models is a very standard
technique~\cite{LavisBell1999}, we would like to emphasize that
our variational-free-energy approach considerably simplifies
the derivation of the recursion equations, which may be of use
in particular when dealing with models of considerable
complexity like the current one. This approach is not fully
original, because it has been demonstrated for instance in
certain specialized literature dealing with spin glasses (where
a \emph{quenched} variational free energy is
considered)~\cite{MezardParisi2001}, but, to the best of our
knowledge, it has never been practically exploited, at least in
the context of polymer
models~\cite{Pretti2002,SerraStilckCavalcantiMachado2004,ZaraPretti2007,OliveiraStilckSerra2008,NetoStilck2013,LiseMaritanPelizzola1998,BuzanoPretti2002,Pretti2006,SerraStilck2007,NetoStilck2008}.


\appendix

\section{Recursion equations}
\label{app:rec}

In this Appendix, we shall report a detailed derivation of the
recursion equations for the Husimi-lattice model.
\begin{table*}[t]
  \setlength{\unitlength}{0.8mm}
  \caption
  {
    The first row reports the possible configurations of the polymer on a lattice site,
    with $w_{i}$ denoting the corresponding cavity fields
    associated to the two branches attached to the site
    (top-right and bottom-left corners).
    The ``prime'' symbol in parentheses denotes the possibility of both fields
    $w_{2^{\protect\vphantom{\prime}}}$ and $w_{2^{\prime}}$.
    The second row reports the contribution of each site
    to the site partition function $z$ (see the text).
  }
  \begin{ruledtabular}
  \begin{tabular}{cccccccccc}
    \begin{picture}(16,16)(-8,-8)
      \linethickness{0.2\unitlength}
      \put( 0, 0){\line( 1, 0){7}}
      \put( 0, 0){\line(-1, 0){7}}
      \put( 0, 0){\line( 0, 1){7}}
      \put( 0, 0){\line( 0,-1){7}}
      \put(-6.5,-5){\makebox[0pt][s]{$w_{0^{\vphantom{(\prime)}}}$}}
      \put( 7.5, 4){\makebox[0pt][r]{$w_{0^{\vphantom{(\prime)}}}$}}
    \end{picture}
    &
    \begin{picture}(18,16)(-9,-8)
      \linethickness{0.2\unitlength}
      \put( 0, 0){\line( 1, 0){7}}
      \put( 0, 0){\line(-1, 0){7}}
      \put( 0, 0){\line( 0, 1){7}}
      \put( 0, 0){\line( 0,-1){7}}
      \linethickness{0.8\unitlength}
      \put(0, 0){\line( 1, 0){7}}
      \put(0, 0){\line(-1, 0){7}}
      \put(-8,-5){\makebox[0pt][s]{$\frac{1}{2}w_{1^{\vphantom{(\prime)}}}$}}
      \put( 9, 4){\makebox[0pt][r]{$\frac{1}{2}w_{1^{\vphantom{(\prime)}}}$}}
    \end{picture}
    &
    \begin{picture}(18,16)(-9,-8)
      \linethickness{0.2\unitlength}
      \put( 0, 0){\line( 1, 0){7}}
      \put( 0, 0){\line(-1, 0){7}}
      \put( 0, 0){\line( 0, 1){7}}
      \put( 0, 0){\line( 0,-1){7}}
      \linethickness{0.8\unitlength}
      \put(0, 0){\line( 0, 1){7}}
      \put(0, 0){\line( 0,-1){7}}
      \put(-8.4,-5){\makebox[0pt][s]{$\frac{1}{2}w_{1^{\vphantom{(\prime)}}}$}}
      \put( 9.4, 4){\makebox[0pt][r]{$\frac{1}{2}w_{1^{\vphantom{(\prime)}}}$}}
    \end{picture}
    &
    \begin{picture}(18,16)(-9,-8)
      \linethickness{0.2\unitlength}
      \put( 0, 0){\line( 1, 0){7}}
      \put( 0, 0){\line(-1, 0){7}}
      \put( 0, 0){\line( 0, 1){7}}
      \put( 0, 0){\line( 0,-1){7}}
      \linethickness{0.8\unitlength}
      \put( 2, 0){\line( 1, 0){5}}
      \put( 0,-2){\line( 0,-1){5}}
      \qbezier( 2, 0)(0,0)( 0,-2)
      \put(-8.4,-5){\makebox[0pt][s]{$\frac{1}{2}w_{1^{\vphantom{(\prime)}}}$}}
      \put( 9, 4){\makebox[0pt][r]{$\frac{1}{2}w_{1^{\vphantom{(\prime)}}}$}}
    \end{picture}
    &
    \begin{picture}(18,16)(-9,-8)
      \linethickness{0.2\unitlength}
      \put( 0, 0){\line( 1, 0){7}}
      \put( 0, 0){\line(-1, 0){7}}
      \put( 0, 0){\line( 0, 1){7}}
      \put( 0, 0){\line( 0,-1){7}}
      \linethickness{0.8\unitlength}
      \put(-2, 0){\line(-1, 0){5}}
      \put( 0, 2){\line( 0, 1){5}}
      \qbezier(-2, 0)(0,0)( 0, 2)
      \put(-8,-5){\makebox[0pt][s]{$\frac{1}{2}w_{1^{\vphantom{(\prime)}}}$}}
      \put( 9.4, 4){\makebox[0pt][r]{$\frac{1}{2}w_{1^{\vphantom{(\prime)}}}$}}
    \end{picture}
    &
    \begin{picture}(16,16)(-8,-8)
      \linethickness{0.2\unitlength}
      \put( 0, 0){\line( 1, 0){7}}
      \put( 0, 0){\line(-1, 0){7}}
      \put( 0, 0){\line( 0, 1){7}}
      \put( 0, 0){\line( 0,-1){7}}
      \linethickness{0.8\unitlength}
      \put( 2, 0){\line( 1, 0){5}}
      \put( 0, 2){\line( 0, 1){5}}
      \qbezier( 2, 0)(0,0)( 0, 2)
      \put(-7,-5){\makebox[0pt][s]{$w_{0^{\prime\vphantom{(\prime)}}}$}}
      \put( 8, 4){\makebox[0pt][r]{$w_{2^{\vphantom{(\prime)}}}$}}
    \end{picture}
    &
    \begin{picture}(16,16)(-8,-8)
      \linethickness{0.2\unitlength}
      \put( 0, 0){\line( 1, 0){7}}
      \put( 0, 0){\line(-1, 0){7}}
      \put( 0, 0){\line( 0, 1){7}}
      \put( 0, 0){\line( 0,-1){7}}
      \linethickness{0.8\unitlength}
      \put(-2, 0){\line(-1, 0){5}}
      \put( 0,-2){\line( 0,-1){5}}
      \qbezier(-2, 0)(0,0)( 0,-2)
      \put(-7,-5){\makebox[0pt][s]{$w_{2^{\vphantom{(\prime)}}}$}}
      \put( 8, 4){\makebox[0pt][r]{$w_{0^{\prime\vphantom{(\prime)}}}$}}
    \end{picture}
    &
    \begin{picture}(16,16)(-8,-8)
      \linethickness{0.8\unitlength}
      \put( 2, 0){\line( 1, 0){5}}
      \put( 0, 2){\line( 0, 1){5}}
      \put(-2, 0){\line(-1, 0){5}}
      \put( 0,-2){\line( 0,-1){5}}
      \qbezier( 2, 0)(0,0)( 0, 2)
      \qbezier(-2, 0)(0,0)( 0,-2)
      \put(-7,-5){\makebox[0pt][s]{$w_{2^{\vphantom{(\prime)}}}$}}
      \put( 8, 4){\makebox[0pt][r]{$w_{2^{\vphantom{(\prime)}}}$}}
    \end{picture}
    &
    \begin{picture}(22,16)(-11,-8)
      \linethickness{0.8\unitlength}
      \put( 2, 0){\line( 1, 0){5}}
      \put( 0, 2){\line( 0, 1){5}}
      \put(-2, 0){\line(-1, 0){5}}
      \put( 0,-2){\line( 0,-1){5}}
      \qbezier( 2, 0)(0,0)( 0,-2)
      \qbezier(-2, 0)(0,0)( 0, 2)
      \put(-9.5,-5){\makebox[0pt][s]{$w_{2^{(\prime)}}$}}
      \put(11, 4){\makebox[0pt][r]{$w_{2^{(\prime)}}$}}
    \end{picture}
    &
    \begin{picture}(22,16)(-11,-8)
      \linethickness{0.8\unitlength}
      \put( 0, 0){\line( 1, 0){7}}
      \put( 0, 0){\line( 0, 1){7}}
      \put( 0, 0){\line(-1, 0){7}}
      \put( 0, 0){\line( 0,-1){7}}
      \put(-9.5,-5){\makebox[0pt][s]{$w_{2^{(\prime)}}$}}
      \put(11, 4){\makebox[0pt][r]{$w_{2^{(\prime)}}$}}
    \end{picture}
    \cr
    $w_{0^{\vphantom{\prime}}}^{2}$
    &
    $\frac{1}{4}w_{1^{\vphantom{\prime}}}^{2}$
    &
    $\frac{1}{4}w_{1^{\vphantom{\prime}}}^{2}$
    &
    $\frac{1}{4}w_{1^{\vphantom{\prime}}}^{2}$
    &
    $\frac{1}{4}w_{1^{\vphantom{\prime}}}^{2}$
    &
    $w_{2^{\vphantom{\prime}}}^{\vphantom{1}} w_{0^{\prime}}^{\vphantom{1}}$
    &
    $w_{2^{\vphantom{\prime}}}^{\vphantom{1}} w_{0^{\prime}}^{\vphantom{1}}$
    &
    $\tau_\mathrm{c} w_{2^{\vphantom{\prime}}}^{2}$
    &
    $\tau_\mathrm{c} \left( w_{2^{\vphantom{\prime}}}^{2}
    + 2 w_{2^{\vphantom{\prime}}}^{\vphantom{1}} w_{2^{\prime}}^{\vphantom{1}} \right)$
    &
    $\tau_\mathrm{x} \left( w_{2^{\vphantom{\prime}}}^{2}
    + 2 w_{2^{\vphantom{\prime}}}^{\vphantom{1}} w_{2^{\prime}}^{\vphantom{1}} \right)$
  \end{tabular}
  \end{ruledtabular}
  \label{tab:site_configurations}
\end{table*}
In Table~\ref{tab:site_configurations} we report all possible
site configurations, along with the cavity fields associated to
the two attached branches. The association rules between fields
and configurations are as follows. First of all, $w_{n}$ is
associated to a configuration with $n$ segments incident from
the branch, for $n=0,1,2$. Moreover, $w_{0}$ and $w_{0'}$
distinguish whether the site is empty or visited, respectively.
Finally, $w_{2'}$ represents the case in which the two incident
segments are connected by a loop whose number of steps does not
diverge in the thermodynamic limit, whereas $w_{2}$ represents
all other cases, namely, either the segments are connected by a
macroscopic loop or they are not connected. Note that a single
segment incident from a branch can arrive from two different
directions. Since in the absence of stiffness we expect the two
possibilities to be equivalent (i.e., the system to be
isotropic), we assume $w_{1}$ to be the total weight of both,
so that a single one has weight ${\frac{1}{2} w_{1}}$.

The second row of Table~\ref{tab:site_configurations} reports
the terms of $z$ corresponding to each configuration. The last
two terms take into account the possibility of a finite loop
($w_{2'}$ field) in one of the two branches, or neither of
them, but not both. In conclusion, the single-site partition
function reads
\begin{equation}
  z = w_{0}^{2} + w_{1}^{2}
      + 2 w_{2}^{\vphantom{1}} \left[ w_{0'}^{\vphantom{1}}
      + (\tau_\mathrm{c} + \tau_\mathrm{x}) w_{2'}^{\vphantom{1}} \right]
      + (2 \tau_\mathrm{c} + \tau_\mathrm{x}) w_{2}^{2}
  \, .
  \label{}
\end{equation}
Let us observe that $w_{0'}$ and $w_{2'}$ appear only in the
linear combination $w_{0'} + (\tau_\mathrm{c} +
\tau_\mathrm{x}) w_{2'}$. We argue that this will be a general
feature of our equations, because this linear combination
represents the total weight of configurations such that the
polymer enters a given branch and exits after a finite number
of steps. In particular, $w_{0'}$ can be regarded as the
zero-step case, in which the polymer visits only the \emph{root
site} of the branch (without a collision), whereas the latter
term includes all other cases. The above observation motivates
the definition of a ``composite'' cavity field, namely,
\begin{equation}
  w_{3} \equiv w_{0'} + (\tau_\mathrm{c} + \tau_\mathrm{x}) w_{2'}
  \, ,
  \label{eq:w3}
\end{equation}
so that the single-site partition function can be rewritten as
\begin{equation}
  z = w_{0}^{2} + w_{1}^{2}
      + 2 w_{2}^{\vphantom{1}} w_{3}^{\vphantom{1}}
      + (2 \tau_\mathrm{c} + \tau_\mathrm{x}) w_{2}^{2}
  \, .
  \label{eq:site_partition_function_(explicit)}
\end{equation}

As far as the cluster partition function is concerned, let us
first observe that we can write
\begin{equation}
  q = \sum_{n=0}^{4} q_{n} \zeta^{n}
  \, ,
  \label{eq:cluster_partition_function_(grand-canonical_vs_canonical)}
\end{equation}
where $\zeta$ is the fugacity and $q_{n}$ can be denoted as the
\emph{canonical} cluster partition function for a cluster with
$n$ segments. According to~\eref{eq:rho}, the segment density
will therefore take the form
\begin{equation}
  \rho = \frac{1}{2q} \sum_{n=0}^{4} n q_{n} \zeta^{n}
  \, .
\end{equation}
In Table~\ref{tab:square_configurations} we report the possible
arrangements of polymer segments on a square cluster, along
with the corresponding cavity fields.
\begin{table*}
  \setlength{\unitlength}{0.8mm}
  \caption
  {
    The first row reports the possible arrangements
    of polymer segments on a square cluster,
    and the corresponding cavity fields.
    The second row reports the number $n$ of polymer segments in each case.
    The third row reports the multiplicity $m$ of each arrangement,
    i.e., the number of different configurations
    that can be generated by rotations.
  }
  \begin{ruledtabular}
  \begin{tabular}{c|cccccc}
    &
    \begin{picture}(30,30)(-8,-8)
      \linethickness{0.2\unitlength}
      \put(-7, 0){\line(1,0){28}}
      \put(-7,14){\line(1,0){28}}
      \put( 0,-7){\line(0,1){28}}
      \put(14,-7){\line(0,1){28}}
      \put(-6,-7){\makebox[0pt][s]{$w_{2}$}}
      \put(-6,-3){\makebox[0pt][s]{$w_{0}$}}
      \put(21,-7){\makebox[0pt][r]{$w_{2}$}}
      \put(21,-3){\makebox[0pt][r]{$w_{0}$}}
      \put(-6,16){\makebox[0pt][s]{$w_{2}$}}
      \put(-6,20){\makebox[0pt][s]{$w_{0}$}}
      \put(21,16){\makebox[0pt][r]{$w_{2}$}}
      \put(21,20){\makebox[0pt][r]{$w_{0}$}}
    \end{picture}
    &
    \begin{picture}(30,30)(-8,-8)
      \linethickness{0.2\unitlength}
      \put(-7, 0){\line(1,0){28}}
      \put(-7,14){\line(1,0){28}}
      \put( 0,-7){\line(0,1){28}}
      \put(14,-7){\line(0,1){28}}
      \linethickness{0.8\unitlength}
      \put( 0, 0){\line(1,0){14}}
      \put(-6,-5){\makebox[0pt][s]{$w_{1}$}}
      \put(21,-5){\makebox[0pt][r]{$w_{1}$}}
      \put(-6,16){\makebox[0pt][s]{$w_{2}$}}
      \put(-6,20){\makebox[0pt][s]{$w_{0}$}}
      \put(21,16){\makebox[0pt][r]{$w_{2}$}}
      \put(21,20){\makebox[0pt][r]{$w_{0}$}}
    \end{picture}
    &
    \begin{picture}(30,30)(-8,-8)
      \linethickness{0.2\unitlength}
      \put(-7, 0){\line(1,0){28}}
      \put(-7,14){\line(1,0){28}}
      \put( 0,-7){\line(0,1){28}}
      \put(14,-7){\line(0,1){28}}
      \linethickness{0.8\unitlength}
      \put( 0, 0){\line(1,0){14}}
      \put( 0, 0){\line(0,1){14}}
      \put(-6,-7){\makebox[0pt][s]{$w_{2}$}}
      \put(-6,-3){\makebox[0pt][s]{$w_{3}$}}
      \put(21,-5){\makebox[0pt][r]{$w_{1}$}}
      \put(-6,18){\makebox[0pt][s]{$w_{1}$}}
      \put(21,16){\makebox[0pt][r]{$w_{2}$}}
      \put(21,20){\makebox[0pt][r]{$w_{0}$}}
    \end{picture}
    &
    \begin{picture}(30,30)(-8,-8)
      \linethickness{0.2\unitlength}
      \put(-7, 0){\line(1,0){28}}
      \put(-7,14){\line(1,0){28}}
      \put( 0,-7){\line(0,1){28}}
      \put(14,-7){\line(0,1){28}}
      \linethickness{0.8\unitlength}
      \put( 0, 0){\line(1,0){14}}
      \put( 0,14){\line(1,0){14}}
      \put(-6,-5){\makebox[0pt][s]{$w_{1}$}}
      \put(21,-5){\makebox[0pt][r]{$w_{1}$}}
      \put(-6,18){\makebox[0pt][s]{$w_{1}$}}
      \put(21,18){\makebox[0pt][r]{$w_{1}$}}
    \end{picture}
    &
    \begin{picture}(30,30)(-8,-8)
      \linethickness{0.2\unitlength}
      \put(-7, 0){\line(1,0){28}}
      \put(-7,14){\line(1,0){28}}
      \put( 0,-7){\line(0,1){28}}
      \put(14,-7){\line(0,1){28}}
      \linethickness{0.8\unitlength}
      \put( 0, 0){\line(1,0){14}}
      \put( 0, 0){\line(0,1){14}}
      \put( 0,14){\line(1,0){14}}
      \put(-6,-7){\makebox[0pt][s]{$w_{2}$}}
      \put(-6,-3){\makebox[0pt][s]{$w_{3}$}}
      \put(21,-5){\makebox[0pt][r]{$w_{1}$}}
      \put(-6,16){\makebox[0pt][s]{$w_{2}$}}
      \put(-6,20){\makebox[0pt][s]{$w_{3}$}}
      \put(21,18){\makebox[0pt][r]{$w_{1}$}}
    \end{picture}
    &
    \begin{picture}(30,30)(-8,-8)
      \linethickness{0.2\unitlength}
      \put(-7, 0){\line(1,0){28}}
      \put(-7,14){\line(1,0){28}}
      \put( 0,-7){\line(0,1){28}}
      \put(14,-7){\line(0,1){28}}
      \linethickness{0.8\unitlength}
      \put( 0, 0){\line(1,0){14}}
      \put( 0, 0){\line(0,1){14}}
      \put( 0,14){\line(1,0){14}}
      \put(14, 0){\line(0,1){14}}
      \put(-6,-7){\makebox[0pt][s]{$w_{2}$}}
      \put(-6,-3){\makebox[0pt][s]{$w_{3}$}}
      \put(21,-7){\makebox[0pt][r]{$w_{2}$}}
      \put(21,-3){\makebox[0pt][r]{$w_{3}$}}
      \put(-6,16){\makebox[0pt][s]{$w_{2}$}}
      \put(-6,20){\makebox[0pt][s]{$w_{3}$}}
      \put(21,16){\makebox[0pt][r]{$w_{2}$}}
      \put(21,20){\makebox[0pt][r]{$w_{3}$}}
    \end{picture}
    \cr
    \hline
    $n$ & $0$ & $1$ & $2$ & $2$ & $3$ & $4$
    \cr
    $m$ & $1$ & $4$ & $4$ & $2$ & $4$ & $1$
  \end{tabular}
  \end{ruledtabular}
  \label{tab:square_configurations}
\end{table*}
We can identify the following rules.

\begin{trivlist}

\item Let $\kappa$ be the number of segments placed on a given
    corner of the square.

\item (i) If ${\kappa = 0}$, then the possible fields are
    $w_{0}$ and $w_{2}$. In the former case the corner site is
    empty. In the latter case the corner site is visited once,
    by a walk along the corner outside the cluster; there can
    occur \emph{contacts} ($\omega$ weight) with other visited sites
    of the cluster.

\item (ii) If ${\kappa = 1}$, then the only possible field is
    $w_{1}$. The site is visited once, by a walk that exits (or
    enters) the cluster; contacts can occur with other visited
    sites.

\item (iii) If ${\kappa = 2}$, then the possible fields are
    $w_{3}$ and $w_{2}$. In the former case the site is visited
    either once, by a walk along the corner inside the cluster,
    or twice, by a walk that exits the cluster and re-enters
    after a finite number of steps. In the latter case the site
    is visited twice. Note that the
    $w_{2}$ field does not incorporate Boltzmann weights of
    crossing or collisions occurring at the corner site,
    whereas the $w_{3}$ field does. The complete field
    combination for the case ${\kappa = 2}$ is therefore ${(2
    \tau_\mathrm{c} + \tau_\mathrm{x}) w_{2} + w_{3}}$.

\end{trivlist}

The above rules, together with
Table~\ref{tab:square_configurations}, explain the following
expressions:
\begin{eqnarray}
  && \!\!\!\!\!
  q_{0}^{\vphantom{1}}
  = w_{0}^{4} + 4 w_{0}^{3} w_{2}^{\vphantom{1}}
  + (4 \omega + 2) w_{0}^{2} w_{2}^{2}
  + 4 \omega^{2} w_{0}^{\vphantom{1}} w_{2}^{3}
  + \omega^{4} w_{2}^{4}
  \, , \nonumber \\ && \!\!\!\!\!
  q_{1}^{\vphantom{1}}
  = 4 \, w_{1}^{2} \left( w_{0}^{2}
  + 2 \omega w_{0}^{\vphantom{1}} w_{2}^{\vphantom{1}}
  + \omega^{3} w_{2}^{2} \right)^{\vphantom{1}}
  \, , \nonumber \\ && \!\!\!\!\!
  q_{2}^{\vphantom{1}}
  = 4 \, w_{1}^{2} \left( w_{0}^{\vphantom{1}}
  + \omega^{2} w_{2}^{\vphantom{1}} \right)
  \left[ (2 \tau_\mathrm{c} + \tau_\mathrm{x}) w_{2}^{\vphantom{1}}
  + w_{3}^{\vphantom{1}} \right]^{\vphantom{1}}
  + 2 \, \omega^{2} w_{1}^{4}
  \, , \nonumber \\ && \!\!\!\!\!
  q_{3}^{\vphantom{1}}
  = 4 \, \omega w_{1}^{2}
  \left[ (2 \tau_\mathrm{c} + \tau_\mathrm{x}) w_{2}^{\vphantom{1}}
  + w_{3}^{\vphantom{1}} \right]^{2}
  \, , \nonumber \\ && \!\!\!\!\!
  q_{4}^{\vphantom{1}}
  = \left[ (2 \tau_\mathrm{c} + \tau_\mathrm{x}) w_{2}^{\vphantom{1}}
  + w_{3}^{\vphantom{1}} \right]^{4}
  - \left( \tau_\mathrm{c} w_{2}^{\vphantom{1}}
  + w_{3}^{\vphantom{1}} \right)^{4}
  \, . \nonumber \\
  \label{eq:cluster_partition_function_(canonical)}
\end{eqnarray}
With respect to the previous discussion, the only extra
ingredient is that we must subtract from the expression of
$q_{4}$ the weight of configurations with finite-length loops
(namely, ${\tau_\mathrm{c} w_{2} + w_{3}}$) on all four
corners. Note that ${\tau_\mathrm{c} w_{2}}$ is the weight
associated to the presence of a walk along the corner inside
the square, \emph{colliding} with a (macroscopic) walk outside.

Let us now also report explicitly the recursion functions
$f_{i}$ appearing in equation~\eref{eq:rec}. Due to the
form~\eref{eq:cluster_partition_function_(grand-canonical_vs_canonical)}
of the cluster partition function, they take the similar form
\begin{equation}
  f_{i} = \sum_{n=0}^{4} f_{i,n} \zeta^{n}
  \, .
  \label{eq:frec}
\end{equation}
The explicit expressions of $f_{i,n}$ turn out to be:
\begin{eqnarray}
  &&
  f_{0,0}^{\vphantom{1}}
  = w_{0}^{3} + 3 w_{0}^{2} w_{2}^{\vphantom{1}}
  + (2 \omega + 1) w_{0}^{\vphantom{1}} w_{2}^{2}
  + \omega^{2} w_{2}^{3}
  \, ,
  \nonumber \\ &&
  f_{0,1}^{\vphantom{1}}
  = 2 w_{1}^{2}
  \left( w_{0}^{\vphantom{1}} + \omega w_{2}^{\vphantom{1}} \right)
  \, ,
  \nonumber \\ &&
  f_{0,2}^{\vphantom{1}} = w_{1}^{2}
  \left[ (2 \tau_\mathrm{c} + \tau_\mathrm{x}) w_{2}^{\vphantom{1}}
  + w_{3}^{\vphantom{1}}\right]
  \, ,
  \nonumber \\ &&
  f_{0,3}^{\vphantom{1}} = 0
  \, ,
  \nonumber \\ &&
  f_{0,4}^{\vphantom{1}} = 0
  \, ; \label{eq:frec-0} \\
  && \nonumber \\
  &&
  f_{1,0}^{\vphantom{1}} = 0
  \, ,
  \nonumber \\ &&
  f_{1,1}^{\vphantom{1}}
  = 2 w_{1}^{\vphantom{1}}
  \left( w_{0}^{2}
  + 2 \omega w_{0}^{\vphantom{1}} w_{2}^{\vphantom{1}}
  + \omega^{3} w_{2}^{2} \right)
  \, ,
  \nonumber \\ &&
  f_{1,2}^{\vphantom{1}}
  = 2 w_{1}^{\vphantom{1}}
  \left( w_{0}^{\vphantom{1}}
  + \omega^{2} w_{2}^{\vphantom{1}} \right)
  \left[ (2 \tau_\mathrm{c} + \tau_\mathrm{x}) w_{2}^{\vphantom{1}}
  + w_{3}^{\vphantom{1}}\right]
  + 2 \omega^{2} w_{1}^{3}
  \, ,
  \nonumber \\ &&
  f_{1,3}^{\vphantom{1}}
  = 2 \omega w_{1}^{\vphantom{1}}
  \left[ (2 \tau_\mathrm{c} + \tau_\mathrm{x}) w_{2}^{\vphantom{1}}
  + w_{3}^{\vphantom{1}}\right]{\!}^{2}
  \, ,
  \nonumber \\ &&
  f_{1,4}^{\vphantom{1}} = 0
  \, ; \label{eq:frec-1} \\
  && \nonumber \\
  &&
  f_{2,0}^{\vphantom{1}} = 0
  \, ,
  \nonumber \\ &&
  f_{2,1}^{\vphantom{1}} = 0
  \, ,
  \nonumber \\ &&
  f_{2,2}^{\vphantom{1}}
  = w_{1}^{2} \left( w_{0}^{\vphantom{1}}
  + \omega^{2} w_{2}^{\vphantom{1}} \right)
  \, ,
  \nonumber \\ &&
  f_{2,3}^{\vphantom{1}}
  = 2 \omega w_{1}^{2}
  \left[ (2 \tau_\mathrm{c} + \tau_\mathrm{x}) w_{2}^{\vphantom{1}}
  + w_{3}^{\vphantom{1}}\right]
  \, ,
  \nonumber \\ &&
  f_{2,4}^{\vphantom{1}}
  = \left[ (2 \tau_\mathrm{c} + \tau_\mathrm{x}) w_{2}^{\vphantom{1}}
  + w_{3}^{\vphantom{1}}\right]{\!}^{3}
  - \left( \tau_\mathrm{c} w_{2}^{\vphantom{1}}
  + w_{3}^{\vphantom{1}}\right){\!}^{3}
  \, ; \label{eq:frec-2} \\
  && \nonumber \\
  &&
  f_{3,0}^{\vphantom{1}}
  = w_{0}^{3}
  + (2 \omega + 1) w_{0}^{2} w_{2}^{\vphantom{1}}
  + 3 \omega^{2} w_{0}^{\vphantom{1}} w_{2}^{2}
  + \omega^{4} w_{2}^{3}
  \, ,
  \nonumber \\ &&
  f_{3,1}^{\vphantom{1}}
  = 2 \omega w_{1}^{2}
  \left( w_{0}^{\vphantom{1}} + \omega^{2} w_{2}^{\vphantom{1}} \right)
  \, ,
  \nonumber \\ &&
  f_{3,2}^{\vphantom{1}}
  = \omega^{2} w_{1}^{2}
  \left[ (2 \tau_\mathrm{c} + \tau_\mathrm{x}) w_{2}^{\vphantom{1}}
  + w_{3}^{\vphantom{1}}\right]
  \, ,
  \nonumber \\ &&
  f_{3,3}^{\vphantom{1}} = 0
  \, ,
  \nonumber \\ &&
  f_{3,4}^{\vphantom{1}}
  = (\tau_\mathrm{c} + \tau_\mathrm{x})
  \left( \tau_\mathrm{c} w_{2}^{\vphantom{1}}
  + w_{3}^{\vphantom{1}} \right){\!}^{3}
  \, . \label{eq:frec-4}
\end{eqnarray}
Let us note that we have obtained the above formulas by a
trivial exercise of derivatives, but, as mentioned in the text,
they can also be interpreted in terms of self-similarity
conditions. Indeed, this is the more usual (recursive) route
followed for this kind of calculations, at least in the context
of polymer models. For this reason we felt it appropriate to
set out also these formulas in full detail.

As mentioned in the text, the numerical technique we employ to
solve the recursion equations is a simple fixed-point method.
Given a tentative set of cavity fields $w_{0},\dots,w_{3}$
(which we shall collectively denote as $w$ in the following
formulas), a new estimate of each field can be computed,
according to equation~\eref{eq:rec}, by an expression of the
form
\begin{equation}
  \hat{w}_{i}(w)
  = \frac{f_{i}(w)}{\sum_{k=0}^{3} f_{k}(w)}
  \, ,
  \label{eq:rec_generica}
\end{equation}
where the denominator takes into account the normalization
condition~\eref{eq:norm}.

A second-order phase transition is characterized by the fact
that a minimum of the free energy becomes a saddle point, so
that the corresponding thermodynamic equilibrium state is no
longer stable. Thermodynamic (in)stability is reflected in the
(in)stability of the fixed point of the recursion equations.
Therefore, in order to determine second-order transitions with
good precision, it is convenient to analyze the eigenvalues of
the Jacobian matrix associated with the recursion equations and
the conditions in which any eigenvalue equals unity. According
to equation~\eref{eq:rec_generica}, the elements of such a
Jacobian matrix can be written as
\begin{equation}
  \frac{\partial \hat{w}_{i}}{\partial w_{j}}(w)
  = \frac{
  \frac{\partial f_{i}}{\partial w_{j}}(w)
  - \hat{w}_{i}(w) \sum_{k=0}^{3} \frac{\partial f_{k}}{\partial w_{j}}(w)
  }{ \sum_{k=0}^{3} f_{k}(w) }
  \, .
\end{equation}
The derivatives of $f_{i}(w)$ can be determined from equations
\eref{eq:frec}--\eref{eq:frec-4}. As far as the transition
lines are concerned, we can locate them numerically as loci of
zeroes of $\det(J-I)$, where $J$ is the Jacobian matrix and $I$
the identity matrix. This can be safely done, because the
peculiar form of the recursion equations allows us to follow a
given solution (fixed point) even in the parameter region where
it becomes unstable. More specifically, if one starts with
tentative fields satisfying ${w_{1} = w_{2} = 0}$ (which
characterizes the 0~phase), the new estimates computed by
equation~\eref{eq:rec_generica} turn out to satisfy rigorously
the same condition. The same holds for the 2~phase,
characterized by the sole condition ${w_{1} = 0}$.


\section{Bethe lattice solution}
\label{app:bethe}

We shall now discuss the solution of the model on the Bethe
lattice. This solution is interesting in particular for the
Wu-Bradley model, because the whole phase diagram, which turns
out to be qualitatively equivalent to that of the Husimi
lattice case, can be worked out analytically, including the
peculiar features of the MC point. Let us remind that the Bethe
lattice can be viewed as a thermodynamic limit of a
random-regular graph, the latter being defined as a random
graph (with ordinary pairwise edges) with fixed coordination
number. Like the Husimi lattice, the Bethe lattice can be
treated locally as if it were a tree-graph, because the length
of closed paths diverges (logarithmically) with the system
size. Still with the idea of approximating the 2d square
lattice model, we choose a coordination number~$4$. Let us warn
the reader that, throughout the following calculations, we
shall use the same symbols used for the Husimi lattice case,
sometimes with a slightly different physical meaning.

The free energy density \emph{per lattice bond} can still be
written in the form
\begin{equation}
  \psi = - \frac{1}{2} \ln q + \ln z
  \, ,
  \label{eq:bethe:free-energy_density}
\end{equation}
where now $q$ is \emph{star-cluster partition function} (where
a star cluster can be defined as a set of lattice bonds
incident to a given site), $z$ is the \emph{single-bond
partition function}, and $1/2$ is the ratio between the number
of star clusters and the number of bonds present in the system.
In analogy to the Husimi lattice solution, the single-bond term
may be viewed as a correction over the cluster term, due to the
overlap between clusters.

\begin{table}
  \setlength{\unitlength}{0.8mm}
  \caption
  {
    Possible configurations of the polymer on a Bethe lattice site:
    $w_{i}$ denote the cavity fields associated to the four branches attached to the site
    (one for each nearest-neighbor).
    The ``prime'' symbol in parentheses denotes the possibility of both fields
    $w_{0^{\protect\vphantom{\prime}}}$ and $w_{0^{\prime}}$.
    The expressions below each figure are the respective
    contributions (including multiplicity) to the star-cluster partition function $q$.
  }
  \begin{ruledtabular}
  \begin{tabular}{ccc}
    \begin{picture}(33,26)(-16,-13)
      \linethickness{0.2\unitlength}
      \put( 0, 0){\line( 1, 0){7}}
      \put( 0, 0){\line( 0, 1){7}}
      \put( 0, 0){\line(-1, 0){7}}
      \put( 0, 0){\line( 0,-1){7}}
      \put(-16,-1){\makebox[0pt][s]{$w_{0^{(\prime)}}$}}
      \put( 17,-1){\makebox[0pt][r]{$w_{0^{(\prime)}}$}}
      \put(2,  9){\makebox[0pt]{$w_{0^{(\prime)}}$}}
      \put(2,-10){\makebox[0pt]{$w_{0^{(\prime)}}$}}
    \end{picture}
    &
    \begin{picture}(27,26)(-13,-13)
      \linethickness{0.2\unitlength}
      \put( 0, 0){\line( 1, 0){7}}
      \put( 0, 0){\line( 0, 1){7}}
      \put( 0, 0){\line(-1, 0){7}}
      \put( 0, 0){\line( 0,-1){7}}
      \linethickness{0.8\unitlength}
      \put(0, 0){\line( 1, 0){7}}
      \put(0, 0){\line(-1, 0){7}}
      \put(-13,-1){\makebox[0pt][s]{$w_{1^{\vphantom{(\prime)}}}$}}
      \put( 14,-1){\makebox[0pt][r]{$w_{1^{\vphantom{(\prime)}}}$}}
      \put(2,  9){\makebox[0pt]{$w_{0^{(\prime)}}$}}
      \put(2,-10){\makebox[0pt]{$w_{0^{(\prime)}}$}}
    \end{picture}
    &
    \begin{picture}(30,26)(-16,-13)
      \linethickness{0.2\unitlength}
      \put( 0, 0){\line( 1, 0){7}}
      \put( 0, 0){\line( 0, 1){7}}
      \put( 0, 0){\line(-1, 0){7}}
      \put( 0, 0){\line( 0,-1){7}}
      \linethickness{0.8\unitlength}
      \put( 2, 0){\line( 1, 0){5}}
      \put( 0, 2){\line( 0, 1){5}}
      \qbezier( 2, 0)(0,0)( 0, 2)
      \put(-16,-1){\makebox[0pt][s]{$w_{0^{(\prime)}}$}}
      \put( 14,-1){\makebox[0pt][r]{$w_{1^{\vphantom{(\prime)}}}$}}
      \put(1,  9){\makebox[0pt]{$w_{1^{\vphantom{(\prime)}}}$}}
      \put(2,-10){\makebox[0pt]{$w_{0^{(\prime)}}$}}
    \end{picture}
    \cr
    $(w_{0^{\vphantom{\prime}}}^{\vphantom{1}} + w_{0^{\prime}}^{\vphantom{1}})^{4}$
    &
    $2 \zeta^{2} w_{1^{\vphantom{\prime}}}^{2}(w_{0^{\vphantom{\prime}}}^{\vphantom{1}} + \omega w_{0^{\prime}}^{\vphantom{1}})^{2}$
    &
    $4 \zeta^{2} w_{1^{\vphantom{\prime}}}^{2}(w_{0^{\vphantom{\prime}}}^{\vphantom{1}} + \omega w_{0^{\prime}}^{\vphantom{1}})^{2}$
    \cr
    \cr
    \begin{picture}(27,26)(-13,-13)
      \linethickness{0.8\unitlength}
      \put( 2, 0){\line( 1, 0){5}}
      \put( 0, 2){\line( 0, 1){5}}
      \put(-2, 0){\line(-1, 0){5}}
      \put( 0,-2){\line( 0,-1){5}}
      \qbezier( 2, 0)(0,0)( 0, 2)
      \qbezier(-2, 0)(0,0)( 0,-2)
      \put(-13,-1){\makebox[0pt][s]{$w_{1^{\vphantom{(\prime)}}}$}}
      \put( 14,-1){\makebox[0pt][r]{$w_{1^{\vphantom{(\prime)}}}$}}
      \put(1,  9){\makebox[0pt]{$w_{1^{\vphantom{(\prime)}}}$}}
      \put(1,-10){\makebox[0pt]{$w_{1^{\vphantom{(\prime)}}}$}}
    \end{picture}
    &
    \begin{picture}(27,26)(-13,-13)
      \linethickness{0.8\unitlength}
      \put( 0, 0){\line( 1, 0){7}}
      \put( 0, 0){\line( 0, 1){7}}
      \put( 0, 0){\line(-1, 0){7}}
      \put( 0, 0){\line( 0,-1){7}}
      \put(-13,-1){\makebox[0pt][s]{$w_{1^{\vphantom{(\prime)}}}$}}
      \put( 14,-1){\makebox[0pt][r]{$w_{1^{\vphantom{(\prime)}}}$}}
      \put(1,  9){\makebox[0pt]{$w_{1^{\vphantom{(\prime)}}}$}}
      \put(1,-10){\makebox[0pt]{$w_{1^{\vphantom{(\prime)}}}$}}
    \end{picture}
    \cr
    $2 \tau_\mathrm{c} \zeta^{4} w_{1^{\vphantom{\prime}}}^{4}$
    &
    $\tau_\mathrm{x} \zeta^{4} w_{1^{\vphantom{\prime}}}^{4}$
  \end{tabular}
  \end{ruledtabular}
  \label{tab:star_configurations}
\end{table}
In Table~\ref{tab:star_configurations} we sketch possible
configurations of the polymer on a lattice site, along with the
cavity fields associated to its nearest neighbors (i.e., the
root sites of the four branches attached to the given site).
Note that each figure represents indeed a multiplicity of
configurations, that can be obtained from one another by
rotations. The cavity fields are defined as follows: $w_{n}$ is
associated to a configuration with ${n=0,1}$ segments incident
from a root site, whereas $w_{0}$ and $w_{0'}$ distinguish
whether the root site is empty or visited, respectively. The
terms of the star-cluster partition function $q$, corresponding
to each configuration, are also reported in
Table~\ref{tab:star_configurations}. Summing these terms, we
obtain
\begin{equation}
  q = (w_{0}^{\vphantom{1}} + w_{0'}^{\vphantom{1}})^{4}
  + 6 \zeta^{2} w_{1}^{2}
  (w_{0}^{\vphantom{1}} + \omega w_{0'}^{\vphantom{1}})^{2}
  + (2 \tau_\mathrm{c} + \tau_\mathrm{x}) \zeta^{4} w_{1}^{4}
  \, .
  \label{eq:bethe:star-cluster_partition_function}
\end{equation}
As far as the single-bond partition function is concerned, we
have
\begin{equation}
  z = w_{0}^{2} + 2 w_{0}^{\vphantom{1}} w_{0'}^{\vphantom{1}} + \omega w_{0'}^{2}
  + \zeta w_{1}^{2}
  \, ,
  \label{eq:bethe:bond_partition_function}
\end{equation}
which can be explained as follows. The first three terms
correspond to an empty lattice bond, where the two neighboring
sites are, respectively, both empty, one visited and one empty,
or both visited. The last term takes into account the case of a
bond visited by a polymer segment, both neighboring sites being
obviously visited as well.

We are now in a position to work out the model solution, since
all the needed information is contained in equations
\eref{eq:bethe:free-energy_density},
\eref{eq:bethe:star-cluster_partition_function},
and~\eref{eq:bethe:bond_partition_function}. Setting at zero
the free-energy derivatives with respect to the cavity fields,
we obtain the self-consistency equations
\begin{eqnarray}
  && w_{0\hphantom{'}}^{\vphantom{1}} \propto
  (w_{0}^{\vphantom{1}} + w_{0'}^{\vphantom{1}})^{3}
  \, , \label{eq:bethe:rec-1} \\
  && w_{0'}^{\vphantom{1}} \propto 3 \zeta^2 w_{1}^{2}
  (w_{0}^{\vphantom{1}} + \omega w_{0'}^{\vphantom{1}})
  \, , \label{eq:bethe:rec-2} \\
  && w_{1\hphantom{'}}^{\vphantom{1}} \propto 3 \zeta w_{1}^{\vphantom{1}}
  (w_{0}^{\vphantom{1}} + \omega w_{0'}^{\vphantom{1}})^{2}
  + (2 \tau_\mathrm{c} + \tau_\mathrm{x}) \zeta^{3} w_{1}^{3}
  \, . \label{eq:bethe:rec-3}
\end{eqnarray}
As previously mentioned, these equations can also be
interpreted as self-similarity conditions, where the
proportionality constant remains undetermined, because of the
invariance of $\psi$ under multiplication of each cavity field
by a constant. Thermal averages can be evaluated as derivatives
of the free energy, and in particular the average number of
segments (segment density) \emph{per bond} reads
\begin{equation}
  \rho = - \frac{\partial \psi}{\partial \ln \zeta}
  \, .
\end{equation}
Exploiting as usual the fact that the derivatives of $\psi$
with respect to the cavity fields vanish at equilibrium points,
we arrive at
\begin{equation}
  \rho = \frac{\zeta w_{1}^{2}}{z}
  \, .
  \label{eq:bethe:rho}
\end{equation}
The last simple expression can also be rationalized from the
previous explanation of the various terms appearing in the
single-bond partition function.

Let us observe that in all the previous equations, the
collision and crossing activities $\tau_\mathrm{c}$ and
$\tau_\mathrm{x}$ always appear only in the combination ${2
\tau_\mathrm{c} + \tau_\mathrm{x}}$. As a consequence, using
the $\tau$ and $\Gamma$ parameters defined by equations
\eref{eq:tau} and~\eref{eq:gamma} (as we shall do in the
following calculations), the model turns out to depend only on
the average~$\tau$, being thus completely unaffected by the
asymmetry parameter~$\Gamma$. This means that the Bethe lattice
model, at odds with the Husimi lattice one, does not exhibit
any ``fine structure'' in the asymmetric case ${\tau_\mathrm{x}
\neq \tau_\mathrm{c}}$.

\subsection{Critical 0-1 transition and TC-1 point}

It can be easily verified that the self-consistency equations
\eref{eq:bethe:rec-1}--\eref{eq:bethe:rec-3} always admit a
solution with ${w_{0'} = w_{1} = 0}$ (and therefore ${\rho =
0}$), whereas $w_{0}$ depends on the normalization condition.
In the following, we shall choose the latter as ${z = 1}$, so
that \eref{eq:bethe:bond_partition_function} implies ${w_{0} =
1}$, whence \eref{eq:bethe:star-cluster_partition_function}
and~\eref{eq:bethe:free-energy_density} trivially lead to
\begin{equation}
  \psi = 0
  \, .
  \label{eq:bethe:free-energy_density_0-phase}
\end{equation}
This solution is obviously identified with the 0~phase.

\begin{figure}
  \resizebox{85mm}{!}{\includegraphics*{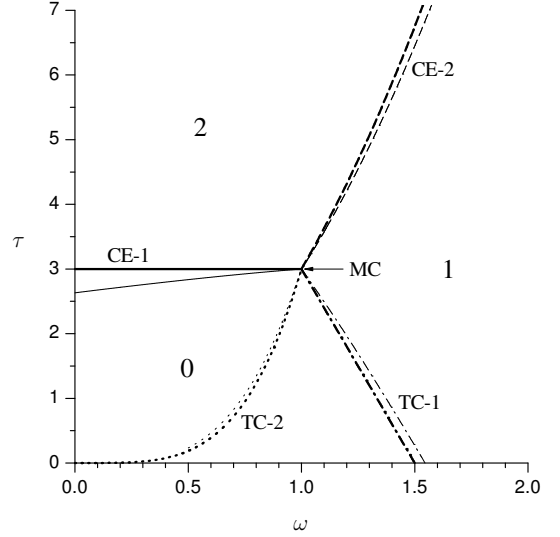}}
  \caption
  {
    Single-chain phase diagram ($\tau$ vs $\omega$)
    for the Bethe lattice model.
    Lines and tags as in Fig.~\ref{fig:omegatau}.
    Thin lines represent the Husimi lattice results
    for the Wu-Bradley model.
  }
  \label{fig:omegatau_bethe}
\end{figure}

In the event of a critical transition (to the 1~phase), we
expect that $w_{0'}$ and $w_{1}$ are small near the transition
line, whereas ${w_{0} \approx 1}$. Let us define the ratios
${\rx \equiv w_{0'}/w_{0}}$ and ${\ry \equiv w_{1}/w_{0}}$.
From \eref{eq:bethe:rec-1}--\eref{eq:bethe:rec-3} we can derive
two self-consistency equations for $\rx$ and $\ry$, namely,
\begin{eqnarray}
  && \rx = \frac{3 \zeta^2 \ry^2 (1 + \omega \rx)}{(1 + \rx)^3}
  \, , \label{eq:bethe:rec-x} \\
  && \ry = \frac{3 \zeta \ry (1 + \omega \rx)^2 + 3 \tau \zeta^3 \ry^3}{(1 + \rx)^3}
  \, , \label{eq:bethe:rec-y}
\end{eqnarray}
which no longer depend on the arbitrary normalization. For
small $\rx$, we have respectively
\begin{eqnarray}
  && \rx = 3 \zeta^2 \ry^2 [ 1 + \mathcal{O}(\rx) ]
  \, , \label{eq:bethe:rec-x_small-x} \\
  && \ry = 3 \zeta \ry [ 1 + (2\omega - 3)\rx + \mathcal{O}(\rx^2) ]
  + 3 \tau \zeta^3 \ry^3 [ 1 + \mathcal{O}(\rx) ]
  \, . \nonumber \\ && \label{eq:bethe:rec-y_small-x}
\end{eqnarray}
Equation~\eref{eq:bethe:rec-x_small-x} confirms that also $\ry$
must be small, and it shows in particular that $\rx =
\mathcal{O}(\ry^2)$. As a consequence, it can also be rewritten
as
\begin{equation}
  \rx = 3 \zeta^2 \ry^2 + \mathcal{O}(\ry^4)
  \, .
\end{equation}
Replacing the latter equation
into~\eref{eq:bethe:rec-y_small-x}, we finally obtain a
self-consistency equation for $\ry$ alone (valid for small
$\ry$), namely
\begin{equation}
  \ry = 3 \zeta \ry + 3 \left[ 3\left(2\omega - 3\right) + \tau \right] \zeta^3 \ry^3 + \mathcal{O}(\ry^5)
  \, .
\end{equation}
The criticality condition (0-1 transition line) occurs when the
coefficient of the degree-1 term on the right-hand side equals
unity, that is for
\begin{equation}
  \zeta = \frac{1}{3}
  \, .
  \label{eq:bethe:0-1_transition}
\end{equation}
Furthermore, the transition changes from second to first order
(TC-1 point) when the degree-3 term vanishes, which happens for
\begin{equation}
  \tau = 9 - 6 \omega
  \, .
\end{equation}
In the single-chain phase diagram $\tau$ vs $\omega$, the
latter equation represents the line of continuous transitions
between the swollen state and the ordinary compact state (TC-1
line). This line turns out to be numerically quite close to the
homologous line determined for the Husimi lattice model (see
Fig.~\ref{fig:omegatau_bethe}).

\subsection{Critical 1-2 transition and TC-2 point}

Equations \eref{eq:bethe:rec-1}--\eref{eq:bethe:rec-3} also
admit a solution with ${w_{0} = w_{0'} = 0}$. With ${z = 1}$,
equations \eref{eq:bethe:rho}
and~\eref{eq:bethe:bond_partition_function} imply ${\rho =
\zeta w_{1}^{2} = 1}$, whence
\eref{eq:bethe:star-cluster_partition_function}
and~\eref{eq:bethe:free-energy_density} lead to
\begin{equation}
  \psi = - {\textstyle \frac{1}{2}} \ln \, ( 3 \tau \zeta^{2} )
  \, .
  \label{eq:bethe:free-energy_density_2-phase}
\end{equation}
This solution can be identified with the 2~phase. Note that a
density of \emph{exactly} $1$ segment per lattice bond means
that \emph{all} lattice sites are doubly visited, i.e., on the
Bethe lattice this is actually a saturated phase, which was not
the case on the Husimi lattice.

In the event of a critical transition (to the 1~phase), we
expect that $w_{0}$ and $w_{0'}$ are small near the transition
line, whereas ${w_{1} \approx 1/\sqrt{\zeta}}$. Let us define
the ratios ${\ra \equiv w_{0}/w_{1}}$ and ${\rb \equiv
w_{0'}/w_{1}}$. As in the previous case, equations
\eref{eq:bethe:rec-1}--\eref{eq:bethe:rec-3} allow us to derive
two recursion equations for $\ra$ and~$\rb$, namely,
\begin{eqnarray}
  && \ra = \frac{(\ra + \rb)^3}{3 \zeta (\ra + \omega \rb)^2 + 3 \tau \zeta^3}
  \, , \label{eq:rec-a} \\
  && \rb = \frac{\zeta (\ra + \omega \rb)}{(\ra + \omega \rb)^2 + \tau \zeta^2}
  \, . \label{eq:rec-b}
\end{eqnarray}
For small $\rb$, from equation~\eref{eq:rec-a} one can easily
argue that ${\ra = \mathcal{O}(\rb^3)}$, which, replaced into
the same equation, leads to
\begin{equation}
  \ra = \frac{1}{3 \tau \zeta^3} \rb^3 + \mathcal{O}(\rb^5)
  \, .
\end{equation}
Plugging this equation into~\eref{eq:rec-b}, we obtain a
self-consistency equation for $\rb$ alone (valid for small
$\rb$), namely,
\begin{equation}
  \rb = \frac{\omega}{\tau \zeta} \rb + \frac{1 - 3 \omega^3 \zeta}{3 \tau^2 \zeta^4} \rb^3 + \mathcal{O}(\rb^5)
  \, .
\end{equation}
The criticality condition (1-2 transition line) occurs when the
coefficient of the degree-1 term on the right-hand side equals
unity, that is for
\begin{equation}
  \zeta = \frac{\omega}{\tau}
  \, .
  \label{eq:bethe:1-2_transition}
\end{equation}
The TC-2 point occurs when we add the condition that the
degree-3 term vanishes:
\begin{equation}
  \zeta = \frac{1}{3 \omega^3}
  \, .
\end{equation}
Putting these conditions together, we obtain
\begin{equation}
  \tau = 3 \omega^4
  \, ,
\end{equation}
which represents the TC-2 line in the single-chain phase
diagram (see Fig.~\ref{fig:omegatau_bethe}). Note that also
this line turns out to be rather close to the analogous one
computed for the Husimi lattice model.

\subsection{First-order 0-2 transition, CE and MC points}

In the previous two subsections, we have argued that the 0
and~2 phases are characterized by constant density values
(${\rho = 0}$ and ${\rho = 1}$, respectively), which implies
that there cannot exist a continuous phase transition between
them. Conversely, a first-order transition exists, and it can
be determined by equating the free energies
\eref{eq:bethe:free-energy_density_0-phase}
and~\eref{eq:bethe:free-energy_density_2-phase}, which yields
\begin{equation}
  \zeta = \frac{1}{\sqrt{3 \tau}}
  \, .
  \label{eq:bethe:0-2_transition}
\end{equation}

The CE-1 point occurs where the 0-2 transition line encounters
the continuous 0-1 transition line (see
Fig.~\ref{fig:omega=0-5}), so that it is defined by the
simultaneous solution of equations
\eref{eq:bethe:0-2_transition}
and~\eref{eq:bethe:0-1_transition}. Eliminating $\zeta$, we
obtain
\begin{equation}
  \tau = 3
  \, ,
  \label{eq:bethe:cep-1_point}
\end{equation}
which is denoted as CE-1 line in Fig.~\ref{fig:omegatau_bethe}.
The CE-2 point occurs where the 0-2 transition line meets the
continuous 1-2 transition line (see Fig.~\ref{fig:omega=1-2}).
This is determined by the simultaneous solution of equations
\eref{eq:bethe:0-2_transition}
and~\eref{eq:bethe:1-2_transition}. Eliminating $\zeta$, we now
obtain
\begin{equation}
  \tau = 3 \omega^2
  \, ,
  \label{eq:bethe:cep-2_point}
\end{equation}
denoted as CE-2 line in Fig.~\ref{fig:omegatau_bethe}. Finally,
the MC point occurs in the event of a simultaneous merging of
the 0-2 line with the (continuous) 0-1 and 1-2 lines (see
Fig.~\ref{fig:omega=1-0_ISAT-model}). As a consequence, it is
defined by the simultaneous solution of equations
\eref{eq:bethe:0-1_transition}, \eref{eq:bethe:1-2_transition},
and \eref{eq:bethe:0-2_transition}, or equivalently of
\eref{eq:bethe:cep-1_point} and~\eref{eq:bethe:cep-2_point},
yielding
\begin{equation}
  \omega = 1 \, , \, \tau   = 3 \, .
  \label{eq:bethe:mp_point}
\end{equation}
It is noticeable that the MC point is the unique transition
point, whose location in the phase diagram remains precisely
the same for both the Husimi and Bethe lattice models, besides
being also equal to the conjectured location of the ISAT
collapse transition for the ordinary 2d square lattice.

\subsection{ISAT model}

\begin{figure}
  \resizebox{85mm}{!}{\includegraphics*{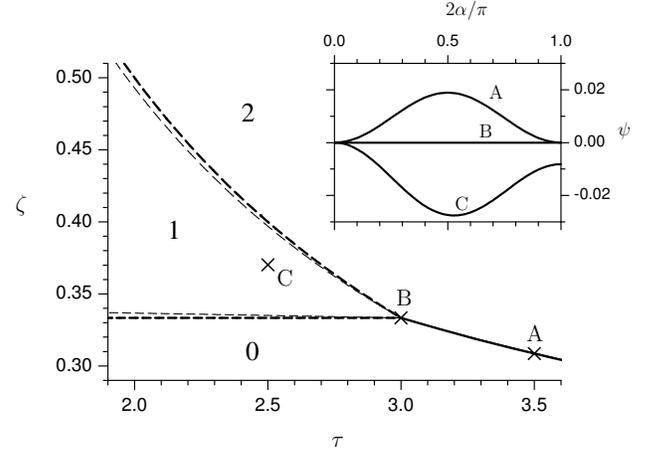}}
  \caption
  {
    Grand-canonical phase diagram ($\zeta$ vs $\tau$)
    for the pure ISAT model (${\omega = 1}$) on the Bethe lattice.
    Lines and phase tags as in Fig.~\ref{fig:omega=1-0_ISAT-model}.
    Thin lines represent the Husimi lattice results.
    The inset displays the variational free-energy density~\eref{eq:bethe:free-energy_density_ISAT}
    as a function of~$\alpha$ (see the text),
    at the points tagged by A, B, C.
  }
  \label{fig:omega=1-0_ISAT-model_bethe}
\end{figure}
Let us now focus on the special case ${\omega = 1}$, which is,
the pure ISAT model. The $\zeta$~vs~$\tau$ phase diagram is
reported in Fig.~\ref{fig:omega=1-0_ISAT-model_bethe}, where we
can observe that the transition lines, determined analytically
by equations \eref{eq:bethe:0-1_transition},
\eref{eq:bethe:1-2_transition},
and~\eref{eq:bethe:0-2_transition}, are numerically very close
to the corresponding ones for the Husimi lattice model. In
particular, as far as the 0-2 transition is concerned, the
numerical discrepancy is so small that it cannot be appreciated
at the scale of the figure.

Let us note that, with ${\omega = 1}$, the variational
free-energy density $\psi$ can be expressed as a function of
only two ``modified'' cavity fields, which we can define as
\begin{eqnarray}
  && u \equiv w_{0} + w_{0'}
  \, , \label{eq:bethe:u-field} \\
  && v \equiv w_{1} \sqrt{\zeta}
  \, . \label{eq:bethe:v-field}
\end{eqnarray}
According to such definitions, the star
\eref{eq:bethe:star-cluster_partition_function} and bond
\eref{eq:bethe:bond_partition_function} partition functions
become respectively
\begin{eqnarray}
  && q = {u}^{4} + 6 \zeta {v}^{2} {u}^{2} + 3 \tau \zeta^{2} {v}^{4}
  \, , \label{eq:bethe:star-cluster_partition_function_uv} \\
  && z = {u}^{2} + {v}^{2}
  \, , \label{eq:bethe:bond_partition_function_uv}
\end{eqnarray}
whereas $\psi$ is still expressed
by~\eref{eq:bethe:free-energy_density}. Due to linearity of
\eref{eq:bethe:u-field} and~\eref{eq:bethe:v-field}, the free
energy is still invariant under multiplication of $u$ and $v$
by a coefficient, so that we can then freely choose the $(u,v)$
vector to have a unit magnitude. As a consequence, it can be
parameterized as function of a single angular coordinate
$\alpha$ as
\begin{equation}
  (u, v) = (\cos \alpha, \sin \alpha)
  \, ,
  \label{eq:bethe:uv-fields}
\end{equation}
with ${\alpha \in [0,\pi/2]}$, because $u$ and $v$ must be both
nonnegative. Replacing equation~\eref{eq:bethe:uv-fields} into
\eref{eq:bethe:star-cluster_partition_function_uv}
and~\eref{eq:bethe:bond_partition_function_uv}, and then
into~\eref{eq:bethe:free-energy_density}, by simple algebra we
arrive at
\begin{equation}
  \psi = - {\textstyle \frac{1}{2}} \ln \left[
  1 + {\textstyle \frac{1}{2}} (3 \zeta - 1) \sin^2(2\alpha)
  + (3 \tau \zeta^2 - 1) \sin^4 \alpha
  \right]
  \, .
  \label{eq:bethe:free-energy_density_ISAT}
\end{equation}
Moreover, taking into account \eref{eq:bethe:v-field}
and~\eref{eq:bethe:rho}, we obtain
\begin{equation}
  \rho = \sin^2 \alpha
  \, .
  \label{eq:bethe:rho_alpha}
\end{equation}
In conclusion, both the free-energy density $\psi$ and the
segment density $\rho$ have been expressed as functions of a
unique variational parameter $\alpha$.

Equilibrium points can now be determined from
equation~\eref{eq:bethe:free-energy_density_ISAT} by setting at
zero the derivative of $\psi$ with respect to $\alpha$. It
turns out that two constant solutions exist, namely, ${\alpha =
0}$ and ${\alpha = \pi/2}$, corresponding respectively to the 0
and 2 phases. Taking also into account
\eref{eq:bethe:rho_alpha}, the third solution, corresponding to
the 1~phase, can be characterized by
\begin{equation}
  \tau = \frac{1}{\zeta} \left[
  1 - \left( 1 - \frac{1}{3 \zeta} \right) \left( \frac{1}{\rho} - 1 \right)
  \right]
  \, .
\end{equation}
The latter equation can be regarded as a family of contour
lines of the density $\rho$ in the $\zeta$ vs $\tau$ diagram.
It is easily observed that all such lines pass through the
MC~point~\eref{eq:bethe:mp_point}, in agreement with the result
obtained numerically for the Husimi lattice model.

We are now also in a position to probe analytically (for the
Bethe case) the peculiar characterization of the MC~point that
has been discussed in the text. Indeed, in
equation~\eref{eq:bethe:free-energy_density_ISAT} we see that
the occurrence of the 0-2 transition
condition~\eref{eq:bethe:0-2_transition} implies that the
${\sin^{4} \alpha}$ term vanishes. Therefore, the free energy
barrier is due entirely to the ${\sin^{2}(2 \alpha)}$ term,
whose amplitude vanishes precisely at the MC~point ${\zeta =
1/3}$. In the inset of
Fig.~\ref{fig:omega=1-0_ISAT-model_bethe}, we have reported
$\psi$ as a function of $\alpha$ at different points in the
$\zeta$~vs~$\tau$ plane, namely, those tagged by A, B, C.
Point~A is located along the 0-2 first-order transition line,
so that we can observe a free-energy barrier between two
equivalent minima. Point~B coincides with the MC~point, so that
we observe a vanishing barrier. Finally, point~C is located in
the region of the 1~phase, so that we can observe the absolute
free-energy minimum at an intermediate density value. Note
that, since the point is located slightly off the analytical
continuation of the 0-2 transition line, the ${\sin^{4}
\alpha}$ term no longer vanishes, so that the free energies of
the (unstable) $0$ and $2$ phases become different.




\end{document}